# Mean radius and shape of Pluto and Charon from *New Horizons* images

Francis Nimmo[1], Orkan Umurhan[2], Carey M. Lisse[3], Carver J. Bierson[1], Tod R. Lauer[4], Marc W. Buie[5], Henry B. Throop[6], Josh A. Kammer[5], James H. Roberts[3], William B. McKinnon[7], Amanda M. Zangari[5], Jeffrey M. Moore[2], S. Alan Stern[5], Leslie A. Young[5], Harold A. Weaver[3], Cathy B. Olkin[5], Kim Ennico[2] and the New Horizons GGI team

## Abstract

Approach images taken by the LORRI imaging system during the *New Horizons* spacecraft encounter have been used to determine the mean radii and shapes of Pluto and Charon. The primary observations are limb locations derived using three independent approaches. The resulting mean radii of Pluto and Charon are 1188.3 ± 1.6 km and 606.0 ± 1.0 km, respectively (2-$\sigma$). The corresponding densities are 1854 ± 11 kg m$^{-3}$ and 1701 ± 33 kg m$^{-3}$ (2-$\sigma$). The Charon radius value is consistent with previous Earth-based occultation estimates. The Pluto radius estimate is consistent with solar occultation measurements performed by the ALICE and Fine Sun Sensor instruments on *New Horizons*. Neither Pluto nor Charon show any evidence for tidal/rotational distortions; upper bounds on the oblateness are <0.6% and <0.5%, respectively.

6 tables; 5 figures; 5 appendices

[1]Dept. Earth and Planetary Sciences, University of California Santa Cruz, Santa Cruz CA 95064, [2]NASA Ames Research Center, Moffett Field, CA 94035, [3]Johns Hopkins University Applied Physics Laboratory, Laurel MD 20723, [4]NOAO, P.O. Box 26732, Tucson, AZ 85726, [5]Southwest Research Institute, 1050 Walnut St. Suite 300, Boulder CO 80302, [6]Planetary Science Institute, 1700 E Fort Lowell

Suite 106, Tucson AZ 85719, [7]Dept. Earth and Planetary Sciences, Washington University, St. Louis MO 63130.

**1. Introduction**

Knowing the mean radius of both Pluto and Charon is a pre-requisite for determining their bulk density, which in turn has implications for their bulk composition and, potentially, their mode of formation. For instance, models in which Charon formed via a giant impact [Canup 2005, Desch 2015] make different predictions from models in which both bodies formed by direct collapse of gravitationally-bound clumps of ``pebbles'' [Nesvorny *et al.* 2010]. Likewise, the long-wavelength shape of these bodies - departures from sphericity, such as rotational flattening – can potentially provide information on their internal structure and evolution [McKinnon and Singer 2014]. This paper provides a preliminary assessment of the mean radius and shape of both Pluto and Charon, based primarily on optical imaging provided by the *New Horizons* spacecraft during its recent flyby [Stern *et al.* 2015].

Prior to the *New Horizons* flyby, radii for Pluto and Charon had been estimated using Earth-based observations of stellar occultations and mutual events (Table 1). For Charon, recent occultation results were internally quite consistent and yielded a radius of around 605 km [Person *et al.* 2006, Sicardy *et al.* 2006], with some outliers being ascribed to local topography. For Pluto, however, the existence of a thin atmosphere [e.g. Binzel and Hubbard 1997, Person *et al.* 2013] increased the uncertainty in radius estimates; for instance, Lellouch *et al.* [2009] reported a range of 1169-1193 km. Mutual event estimates were not affected by the atmosphere, but were thought to be affected by orbital uncertainties, limb darkening and albedo variations. Young and Binzel [1994] derived radii for Pluto and Charon of $1179\pm 24$ km and $629\pm21$ km, respectively. Buie *et al.* [1992] reported values for Pluto and Charon of $1150\pm7$ km and $593\pm10$ km, respectively; by using other techniques (stellar occultations and Hubble images, respectively) to fix the radius and orbit of Charon, the radius of Pluto was later updated to 1189.5 km by Tholen [2014].

Because of these uncertainties the densities of Pluto and Charon were barely distinguishable at the 2-σ level [e.g. Brozovic *et al.* 2015]. As initially reported in Stern *et al.* [2015] and elucidated in more detail here, one of the major results of the *New Horizons* mission is that Pluto is definitely denser than Charon, by about 9%.

Pluto and Charon today occupy a doubly-synchronous state [Dobrovolskis *et al.* 1997, Buie *et al.* 1997]. If they respond to the present-day tidal and rotational potentials as fluid bodies, then Pluto will closely approximate an oblate spheroid, while Charon will be triaxial. For Pluto, the flattening is approximated by $R\omega^2 h_2/2g$ [e.g. Murray and Dermott 2000; also see Section 5 below]. Here $R$ is the radius, $\omega$ the rotation angular frequency, $h_2$ a dimensionless constant of order unity for a fluid, and $g$ the surface gravity. Deviations from sphericity for a present-day, fluid Pluto are thus expected to be ~0.05%, with similar deviations expected at Charon. In principle, these deviations can be used to determine a body's moment of inertia (as long as it is behaving in a fluid-like fashion); such techniques have been used with some success for the moons of Jupiter and Saturn [e.g. Dermott & Thomas 1988, Oberst & Schuster 2004, Thomas *et al.* 1998, 2007].

However, earlier in their history, Pluto and Charon were likely closer together and their spin rates were correspondingly faster [Dobrovolskis *et al.* 1997]. Thus, in principle Pluto and/or Charon could have "frozen in" rotational and/or tidal bulges at an early epoch in their history and maintained such "fossil" bulges to the present day. That is, they could resemble the Earth's Moon [e.g. Garrick-Bethell *et al.* 2014] or Iapetus [Castillo-Rogez *et al.* 2007], both of which possess large fossil bulges inconsistent with their present-day spin rates. Whether or not a body can maintain a fossil bulge depends largely on its thermal evolution: bodies that are warm and deformable, or have weak lithospheres or a subsurface ocean, will not maintain such a bulge, while cold and rigid bodies can. Thus, the presence or absence of a fossil bulge at Pluto or Charon provides a constraint on these bodies' internal structure and thermal and orbital evolution [Robuchon and Nimmo 2011, McKinnon and Singer 2014].

The remainder of this paper is organized as follows. We focus primarily on results derived from images taken by *New Horizons*, though we do make limited use of constraints from other techniques. Section 2 describes the primary observation set used to determine radius and shape, while Section 3 describes the techniques used to convert a single image into an estimate of radius and shape. This section includes a preliminary analysis of likely uncertainties by analyzing synthetic data sets. Section 4 extends the approach outlined in Section 3 to fitting results from multiple images simultaneously, while Section 5 summarizes our results and discusses the implications. The Appendices provide further details of specific aspects of our analysis.

## 2. Observations

### 2.1 LORRI

The primary instrument used in this analysis was the Long-Range Reconnaissance Imager (LORRI), the characteristics of which are described in Cheng *et al.* [2008]. In brief, LORRI comprises a 1024x1024 pixel CCD positioned at the Cassegrain focus of a 20.8 cm Ritchey-Chrétien telescope. Images are obtained in white-light over a 350 – 850 nm bandpass. The mechanical design of the camera ensures high-stability of the optics within the spacecraft environment. Significantly, the camera is operated without a mechanical shutter. Some charge does accumulate during the short frame-transfer time at readout, which is subtracted as part of standard image reduction.

A key parameter in the following analysis is the angular width of a LORRI pixel, because this is required to convert spacecraft range to a length-scale at the image. Based on images of NGC 3532 taken in 2013, the value adopted in this work is 4.963571 +/- 0.000038 μrad/pixel, an update from the value previously reported in Cheng et al. [2008]. The small distortions introduced by the LORRI optics have been quantified [Owen and O'Connell 2011] and their effect is discussed in more detail below and in Appendix E. The point-spread function (PSF) of LORRI is not axisymmetric [Noble *et al.* 2009], but analysis of synthetic images including a realistic PSF suggests that shape determination is negligibly affected by this particular effect (Appendix A).

During the approach phase, LORRI took a series of images of Pluto and Charon at increasingly higher resolution. Our analysis was based on all approach images downlinked as of August 2015. In this work we have only used single-frame images, because fitting a substantial fraction of the illuminated limb is required in order to establish the centre of figure and to search for any oblateness. Mosaics of Pluto and Charon at higher resolutions do exist, but were not used in this study. High-phase departure ("lookback") images of Pluto reveal a pronounced atmospheric haze [Stern *et al.* 2015] which makes identification of the solid surface challenging; those images were not used in the current analysis.

Table 2 presents details of the images used. The highest resolution images are at about 3.7 km/pixel and 2.3 km/pixel for Pluto and Charon, respectively. Among the images used there is a range in resolution of about a factor of 3 (for Pluto) and about 4 (for Charon); adding earlier images would increase the range of longitudes covered, but at the expense of even poorer resolution. The tabulated geometric parameters, discussed in more detail below, are based on the *15sci_rhr* (od122) reconstructed kernel (Steffl et al. 2016) and were derived by the Navigation team using optical navigation and radiometric data following the encounter (see e.g. Jackman et al. 2016, Pelletier et al. 2016). For convenience (and because some images actually combine multiple exposures, see below) we refer to images as "visitXX" with a subscript denoting Pluto or Charon. LORRI images are uniquely identified by their clock time (MET); thus, visit73p can also be referred to as LOR_0299124574 (see Tables 2 & 3).

### 2.2 Combining multiple images

Except near closest approach, the sequencing resulted in two or more images being taken at almost identical epochs. The small, random drift in spacecraft pointing between exposures effectively "dithers" the images in a sequence. This dithering can be used to combine the individual images to produce a single summed and interlaced image that is Nyquist-sampled. This procedure removes the deleterious geometrical effects of aliasing, as well as making optimal use of the spatial information in a given dataset. These images are identified with an asterisk in Table 2.

In detail, the Nyquist-sampled images were produced by the Fourier-based algorithm presented by Lauer [1999]. The algorithm calculates a complex linear combination of the images in Fourier space that uses the different sampling phases to algebraically eliminate the higher-order "satellites" that generate aliasing in a discretely-sampled image with an insufficient sampling frequency. The algorithm does not use any interpolation kernels, which are likely to degrade the spatial resolution of the images. For LORRI, two images dithered with a $1/\sqrt{2}$ pixel-offset in *x* and *y* is sufficient to achieve Nyquist-sampling with a simple diagonal-interlace of the two images. In practice, most of our analysis is done on super-images with pixels twice as fine as the native LORRI scale. Although mis-registration during the combination of images could in principle lead to errors in the derived radius, comparison of individual exposures with the Nyquist-sampled images implied an effect of <0.1 pixels.

When combining images, it is important to use an accurate PSF. A library of PSFs as a function of position over the LORRI field-of-view was generated from extensive observations of star-clusters obtained during the long cruise to Pluto. A given PSF at a particular field location was generated from typically a dozen bright stars falling within a small radius about the location. The image combination again was done with the Lauer

[1999] algorithm, generating a Nyquist-sampled PSF with twice the native LORRI sampling.

### 2.3 Occultations

An independent constraint on the radius of Pluto and Charon may be obtained via occultations. Ground-based occultation results were discussed above; *New Horizons* also obtained three different kinds of occultation measurement during the encounter. The first kind, provided by the radio science experiment (REX) yielded a Pluto radius of 1189.9±0.4 km (Gladstone *et al.,* 2016). The second kind is solar occultations detected using the fine sun sensor (FSS). Details of this analysis are presented in Appendix C. The third kind also uses solar occultations with the ALICE ultraviolet spectrograph. Details of this analysis are presented in Appendix D. Comparison of the occultation results with the LORRI imaging results detailed below is presented in Section 5. All occultation measurements suffer from the drawback that only a single chord across the body is measured, so that local topography, global flattening and/or uncertainties in the chord location can significantly affect the derived radius. As will be seen below, for Pluto all the *New Horizons* occultation data are approximately consistent with the LORRI results, while for Charon significant discrepancies remain. These are most likely due to uncertainties in the chord location.

## 3. Methods
### 3.1. Limb picks

To determine planetary shape from a single image, the first task is to identify points ("limb picks") along the edge of the imaged disk illuminated by sunlight. This is in principle quite straightforward and has been performed successfully for decades [e.g. Dermott and Thomas 1988]. As part of the *New Horizons* encounter effort, three different groups used independent approaches to this problem, each of which is described in more detail below. We then describe how limb picks for an individual image may be used to determine the best-fit radius *R*. This approach is suitable for determining the mean (spherical) shape. To look for departures from sphericity, multiple images must be used, as described in Section 4 below

### 3.1.1 Method A.

The first two methods use a thresholding approach which is very similar to that described in Dermott and Thomas [1988]. In Method A, a series of horizontal and vertical pixel scans across the image are used. Each scan generates brightness as a function of distance along profile. A global value of the background (off-body) brightness $B_{off}$ is calculated based on an initial guess at the body centre and radius. The on-body brightness $B_{on}$ is calculated for each profile by averaging the brightness over $0.5d$ to $0.9d$, where $d$ is the on-body profile length from centre to body edge. The edge location is then taken to be the outermost point at which the brightness equals $B_{off} + f_{thresh}(B_{on}-B_{off})$, where $f_{thresh}$ is an adjustable threshold parameter and cubic spline interpolation is used to obtain sub-pixel accuracy. This approach should be relatively insensitive to effects caused by albedo variations, although in practice these still create some problems (see below). Further details of the technique may be found in Thomason and Nimmo [2015].

This method was calibrated against limb picks for 4 Rhea images taken by the Cassini ISS imaging system [Thomas *et al.* 2010; Thomas, pers. comm.]. For these images the median distance between our individual picks and those of Thomas ranged from 0.13-0.25 pixels, indicating a good level of agreement. Fitting the Thomas and Method A picks for an individual image resulted in absolute differences in the derived radius of 0-0.8 pixels, with a mean absolute value of 0.4 pixels.

### 3.1.2 Method B

Method B is similar in outline to method A. One difference is that an initial set of 50-70 edge picks for the illuminated hemisphere is carried out manually. These initial picks are then used to establish an initial elliptical fit using the least-squares approach of FitzGibbon *et al.* [1999]. This initial ellipse is then used to establish a series of radial transects through each pixel on the illuminated limb; these transects are then used in edge identification. Each transect generates a radial DN profile by cubic spline interpolation. For each transect, points within 3-5% of the nominal edge position are used to determine the maximum and minimum brightness $B_{max}$ and $B_{min}$. The edge is taken to be at a brightness $B_{min} + f_{thresh}(B_{max}-B_{min})$, with the exact position being determined by least-squares fitting or spline interpolation. Obvious outliers were rejected manually.

For a body with a simple phase curve, one would expect $f_{thresh}\sim50\%$ to produce an accurate pick with either method. For the Rhea images used in method A, $f_{thresh}=50\%$ yielded the best agreement with the Thomas picks, and this value of $f_{thresh}$ was adopted for Charon images. Pluto's phase behavior, however, appears to be more complicated [Buie *et al.* 2010] and it also exhibits extreme albedo variations [Stern *et al.* 2015]. By generating synthetic Plutos with a range of phase curves and realistic large-scale albedo variations, it was found that for Method A $f_{thresh}$ in the range 30-40% generally produced the most reliable results, while for Method B $f_{thresh}$=50% was appropriate (Appendix A). A lower threshold generally results in a larger radius.

### 3.1.3 Method C

Method C, also termed the Maximum Derivative Method, uses the 2-D spatial derivative filters available in IDL (e.g. the Roberts, Sobel, and Prewitt filters) to create a gradient filtered image of Pluto or Charon. Except in the case of a high contrast feature on the body's surface, in any row or column of a LORRI image the maximal gradient should lie at the boundary between the body and interplanetary space.

Two different approaches can then be used to solve for the size and oblateness of Pluto and Charon. In the first approach, the *x-y* coordinates of the MDM image pixels are transformed into *r-θ* coordinates, in which a spherical body would have an edge at *r* = constant i.e. a horizontal line. The best-fit radius is then found by carrying out a series of horizontal scans and finding the scan that minimizes the RMS misfit between the interpolated MDM pixel values and a constant model value. By empirical testing, a constant value one-third of the maximum MDM pixel value was found to produce the best and quickest convergence. Only illuminated azimuths are included in the misfit calculation. This process is repeated for different values of the center coordinates $x_0,y_0$ to find the best-fit $r,x_0$ and $y_0$. The method also allowed an initial search for deviations from spherical symmetry to be accomplished, by looking at the deviations in the data from the straight-line (*r* = constant) solution. No such deviations were found; further discussion of this topic is deferred to Section 4.

In the second approach, the location of the maximum gradient is determined for each half-row and half-column of the image centered above and below the center of brightness of the image. This approach produces a set of (*x,y*) picks that can then be analyzed in the same way as the other two approaches above.

The MDM technique was calibrated using LORRI images [1] of the large airless Galilean moons Ganymede and Callisto during the *New Horizons* flyby of Jupiter in February 2007 [Grundy *et al.* 2007], for which the mean radii are well known [Davies *et al.* 1998, Anderson *et al.* 2001], and by using synthetic model images of Pluto created by M. Buie (Appendix A). In both cases the MDM technique was able to reproduce the size of the objects to an accuracy of ~0.5 pixel, similar to that for method A.

---

[1] Images used were: (Ganymede) LOR_0034876394, LOR_0034784234, (Callisto) LOR_0034731614, LOR_0034939034, lor_0034858514

The advantage of the MDM technique is that there are no tunable parameters, unlike Methods A and B. On the other hand, the gradient technique tends to introduce additional noise, which produces more scatter in picks generated by this technique. As with the other methods, this technique is adversely affected when the limb structure is poorly defined, or high contrast structures that are off- limb dominate a region of the image.

**3.2 Radius fits for single images**

In general, only a subset of the edge points identified by the algorithms will actually be on the illuminated portion (i.e. the limb). To calculate which points are actually illuminated, the following procedure is adopted. Any individual point ($x,y$) may be converted to an equivalent latitude, longitude position ($\phi,\theta$) on a spherical body given the image coordinates of the body's centre ($x_0,y_0$), the latitude and longitude of the centre ($\phi_0,\theta_0$), the body's radius $R$ and the orientation of the rotation pole relative to the image ($\varphi$). Here this conversion is carried out using an orthographic projection. This approximation is appropriate for the distant images used here; for instance, the highest-resolution Pluto image (visit75p) took place at a distance to Pluto's centre of 736,064 km, or 619 Pluto radii. The rotation pole is assumed normal to the orbit plane (zero obliquity) and the Charon-Pluto line sets the location of the prime meridians. It is then straightforward to calculate the angular distance between the point of interest and the subsolar point ($\phi_s,\theta_s$); if this distance is <90°, then the point is illuminated. The geometric parameters for each image are listed in Table 2.

The main complication with this procedure is that it is iterative. An initial guess at the parameters $x_0,y_0$ and $R$ will yield an initial set of illuminated limb points. These may then be fit to some model shape (see below) which will yield updated parameter values, and thus a new set of illuminated points, which then must be fit. In practice, the parameter sensitivity is not very strong, so the iteration procedure was carried out manually, and typically converged in 2-3 iterations.

Figure 1 shows limb picks obtained from a series of images for Pluto and Charon using method A, where red points are illuminated and black points are not. Fig 1c shows a clear example of where the algorithm fails to pick the image edge of the non-illuminated hemisphere due to large albedo variations. This sometimes also occurs on the illuminated hemisphere, in which case the erroneous picks had to be removed manually. The gaps in Figs 1a and 1d are due to this manual removal process.

Given illuminated limb picks for an individual image, the best-fit ellipse or circle defined by those picks can be established. The equation for the ellipse is

$$\left(\frac{x-x_0}{a}\right)^2 + \left(\frac{y-y_0}{b}\right)^2 = 1 \qquad (1)$$

where the fitting parameters to be determined are the semi-major and minor axes $a,b$ and the centre of the figure $x_0,y_0$. Note that here it is assumed that the shape axes coincide with the image axes. In practice, as will be shown below, fitting picks from a single image is only suitable for determining the mean (circular) radius $R$.

Although many analytical techniques are available to fit circles or ellipses to data points [e.g. FitzGibbon *et al.* 1999] here we elected to use a brute-force grid-search method with a step size of 0.1 pixels in $x_0,y_0$, $a$ and $b$. The advantage of doing so is that it permits establishment of the uncertainty ellipse. This process is iterative, because the illuminated points depend on the fitting parameters chosen (see above).

In the calculations below for limb-pick methods the RMS misfit $\chi$ is used as a metric of fit, and is defined as

$$\chi = \sqrt{\frac{1}{N}\sum_N (r_i - r_i')^2} \qquad (2)$$

where $r_i$ and $r'_i$ are the actual and model radial distance from the centre of figure to the point in question, and $N$ is the total number of points.

### 3.3 Sources and quantification of uncertainty

It is obviously important to assess the level of uncertainty in derived quantities. In this preliminary analysis we have identified five main sources of uncertainty: image smear, camera distortion, rough limb topography, errors in centroid location, and the PSF. We discuss each of these below.

Image smear arises because small angular rotation of the spacecraft during a single exposure results in motion of the image plane relative to the target. The drift rate can be computed *a posteriori*, and total offsets during exposure are tabulated in Table 3 for each image. The offset is typically a few tenths of a pixel, of order 1 km. Fortunately, below we show that comparison of the two pairs of high-resolution images – having smear that differs in magnitude and direction - suggest that the smear has a minimal effect on radius determination. For visit73p and visit75p the radii are 1189.7 and 1188.8 km using method A, while for visit76c and visit74c the derived radii are 605.7 and 605.6 km. The small effect of image smear makes intuitive sense: smear only acts in one direction, so that the net effect on radius determination will be at most half the total smear amount. In principle, smear might have more of an effect on ellipse shape determination, but a similar analysis in Section 4.2 suggests that smear does not unduly affect ellipse determination either.

The LORRI optics produce a small level of "pin-cushion" distortion over the focal plane. This is strongest at the edges of the CCD array [Owen & O'Connell 2011] and is really only of importance when the imaging target nearly fills the frame. Appendix E provides a model description of the distortion. Correction of this distortion for the two full-frame Pluto images results in a reduction in radius of 0.1-0.2 pixels, or 0.35-0.7 km (see Table 4 below); for Charon full-frame images the reduction is 0.1 pixels, or 0.2 km. In the following analysis, except when noted otherwise, picks from the two full-frame Pluto images only were corrected for distortion prior to analysis.

To assess the effects of rough limb topography and centroid determination we have made use of synthetic images as described in Appendix B. Briefly, a set of synthetic limb picks are derived, including contributions from an elliptical shape, rough topography and random noise. These picks are fit using exactly the same technique as the real picks (Section 3.2). By fitting many sets of synthetic picks, the RMS misfit value associated with a particular probability of including the true value (e.g. 68%=1-$\sigma$) can be determined. For a single image the 1-$\sigma$ misfit value was found to be 1.022$\chi_{min}$, where $\chi_{min}$ is the best-fit RMS; in what follows we took the 2-$\sigma$ value to be 1.044$\chi_{min}$ and used this value of $\chi$ to determine our 2-$\sigma$ error ellipses. For combined images the analysis is somewhat more complicated, but as discussed in Appendix B we estimate that the 2-$\sigma$ error ellipse in this case is given by $\chi \approx 1.1\chi_{min}$ with typical uncertainties in $R$, $a$ and $c$ of 1,1 and 2 km for Pluto.

For the $r,\theta$ ("unwrap") version of Method C the 2-$\sigma$ value for individual images was taken to be $\chi = \chi_{min}(1 + 2.357/\sqrt{N})$ where $N$ is the number of data points. This approach yields comparable uncertainties in radius (Tables 4,5).

The synthetic images described in Appendix A included a realistic LORRI PSF. No systematic effects of this PSF on the derived radii were found, leading us to believe that the effect of the PSF on our results is negligible.

Limb profiles can result in radii that are biased high because depressions (e.g. craters) are masked by elevated topography in front or behind [e.g. Dermott & Thomas 1988]. Given an estimate of the variation in roughness with wavelength, this effect can be quantified [Nimmo *et al.* 2010], but the effect is usually very small and has not been included here. Elevated topography can also cause the ground location of the limb point to vary from the nominal position by an amount $\approx \sqrt{2h/R}$ where $h$ is the topography, up to about 5° for Pluto. Such deviations are unlikely to affect the global shape determination significantly.

**4 Results**

**4.1 Results for single images**

Figure 2 shows the results of fitting individual images with either circles or ellipses. Figures 2a and 2d show the misfit as a function of $x_0$ and $y_0$ for a circular model; for each ($x_0,y_0$) pair, the local best-fit radius was used. The formal 2σ uncertainties are 1-2 pixels (see below), but the important point is that in both cases the error ellipses are elongated – in the *y*-direction for visit73p and in the *x*-direction for visit74c. Inspection of Figs 1d and 1i reveals the reason. For visit73p, the distribution of the limb picks means that the shape is poorly constrained in the up-down (*y*) direction, while for visit74c, the different distribution of picks means that the left-right (*x*) direction is poorly constrained.

The effect of this uncertainty only has a small effect on the best-fit radius determination (0.2 pixels) and is thus not critical if only *R* is to be determined. However, the same effects make accurate fitting of an ellipse much more challenging. Figs 2b and 2d show the misfit as a function of the ellipse axes *a* and *b*. In both cases, one of these axes is very poorly determined and in Fig 2e the apparent best-fit shape departs significantly from sphericity. The reason is seen in Figs 2c and 2f: as explained above, the centroid location in the *y* and *x* directions, respectively, are not well-constrained. Because of the tradeoff between *a* and $x_0$, or *b* and $y_0$, investigation of subtle departures from sphericity cannot be carried out with a single image. A different, combined approach is required, which is described in more detail below.

Tables 4 and 5 tabulate the results of fitting circles to individual images, for Pluto and Charon respectively. Several points are of interest. First, as expected, for methods A and B a threshold of 40% for Pluto usually yields a slightly smaller radius than the 30% threshold. The exception is 75p, because of manual removal of erroneous picks in the 40% case. Second, the agreement between methods A and B – using different values of $f_{thresh}$, based on the synthetic images – is very good. The *r-θ* version of method C yields radii with comparable uncertainties but typically very slightly smaller values than the A and B methods, consistent with the synthetic results (Appendix A). The limb-pick version of method C yields large uncertainties, because of the scatter introduced by taking the gradient.

A preliminary assessment of the likely radii of Pluto and Charon can be derived by simply taking a naïve average of the tabulated radius estimates for the two highest resolution images in each case. For Pluto, this yields 1189.3 km; the corresponding standard deviation of the estimates ($N$=7) is 1.2 km (Table 4); for Charon, 606.3 km (Table 5) with a standard deviation of 0.6 km. These estimates obviously ignore the information contained within the other, lower-resolution images, and do not provide any information on deviations from sphericity. As discussed above, single images will not necessarily provide accurate elliptical shapes. The best solution is to use multiple images taken from different orientations; in this way the degeneracies associated with a single image can be avoided. This approach is the subject of Section 4.2 below.

### 4.2 Results for combined images

Section 3.2 outlined how one or several sets of limb picks can be projected onto the surface of a body. Figures 3a and 3b show the results of doing so for Pluto and Charon for the limb picks shown in Fig 1. (We do not plot visit74p, visit75p and visit76c because they lie on top of visit73p and visit74c, respectively - see Table 2). The colors denote the topography, $r_i$-$R$, where $R$ is the best-fit radius and $r_i$ is the distance from the point in question to the centre of the body derived from a particular image. At Pluto the difficulty of identifying the limb in the low albedo Cthulhu Regio (all placenames used in this work are informal) is evident for visit73p (see also Fig 1d). Visit71p crosses the center of Sputnik Planum and identifies it as a topographic low, in agreement with stereo analysis [Moore *et al.*, 2016]. For Charon, Serenity Chasma is perhaps faintly visible in visits 70c,71c and 72c as a local depression. Figs 3c and 3d plot the topography as a function of longitude. The main message is that Pluto and especially Charon exhibit considerable relief, at least ±5 km. Available stereo topography [Moore et al., 2016] confirms that both bodies are indeed topographically rough.

In principle, the calculated topography relative to a sphere (Figs 3a,3b) might be due in part to the underlying shape actually being oblate or ellipsoidal. We can therefore use the same limb picks to explore whether an oblate or ellipsoidal shape provides a significantly better fit, as follows. The radial distance $r'$ to a point at longitude $\phi$, colatitude $\theta$ on the surface of a triaxial ellipsoid with axes $a,b,c$ is given by

$$\frac{1}{r'^2} = \left(\frac{\cos\phi \sin\theta}{a}\right)^2 + \left(\frac{\sin\phi \sin\theta}{b}\right)^2 + \left(\frac{\cos\theta}{c}\right)^2 \qquad (3)$$

For an oblate spheroid, the same result can be used except with $b=a$, while for a tidally-distorted body we expect *(a-c)=4(b-c)* (see below).

Given a set of projected limb picks (see above), the misfit defined by equation (2) can be determined for different combinations of *a*, *b* and *c*. Here we weight the misfit of each image by the inverse of the image resolution; changing the weighting to uniform does not result in appreciable changes to the answers. Note that this approach ignores the effect of any errors in the derived centroid ($x_0, y_0$) for each image; we discuss the effect of this uncertainty on the overall error budget below.

Fig 4a shows the relative misfit as a function of *a* and *c* assuming an oblate Pluto. The uncertainty in *c* is larger than in *a*, primarily because of the paucity of points at high latitudes (Fig 3a). The yellow star denotes the best-fit value, the slanting lines indicate flattening *f=(a-c)/a* of 0.5% and the red contour is a 2-σ error ellipse (Section 3.3). A spherical Pluto is completely consistent with the observations, and the absolute flattening *|f|* does not exceed 0.4% at the 2-σ level. In this plot the highest-resolution image has been corrected for camera distortion (Section 3.3), but neglecting to do so changes the bound on the flattening by only 0.1%. Fitting the same data with a triaxial ellipsoid gives best-fit values of *a*=1188.6 km, *b*=1189.2 km and *c*=1189.4 km but the uncertainties are large and the RMS misfit of 2.71 km does not represent a significant improvement over either the spherical or the oblate fits (Table 6).

Figs 4b and 4c show similar results for Charon. In Fig 4b Charon is assumed to be oblate, while in Fig 4c it is assumed to adopt a hydrostatic ellipsoidal shape, in which *(a-c)=4(b-c)*. The best-fit results show a negative flattening (i.e. *c>a*) which is dynamically implausible and should be taken as an indication of the uncertainties present. However, both plots are consistent with a spherical Charon at the 2-σ level. Allowable (positive) values of *f* do not exceed 0.4% (Fig 4b) and 0.9% (Fig 4c) at the 2-σ level.

The same methodology can be used to deduce the best-fit spherical radius for the two bodies. For Pluto with picks from method A, we find a radius of 1189.0 ± 1.0 km and for Charon 605.5 ± 1.0 km. These estimates are identical within error to the mean values derived from individual high-resolution images in Section 4.1 (1189.3±1.2 and 606.3±0.6 km, respectively).

Table 6 summarizes the best-fit spherical and oblate shapes derived from different techniques and combinations of images. Note that although for each body there are two or three very high resolution images, we only ever use one in a particular solution, as these images were taken at essentially the same epoch. There is very good agreement for the best-fit spherical radii of Pluto and Charon, and there is no evidence within error for any flattening of either body.

The uncertainties given in Table 6 take into account the effects of topographic roughness and uncertainty in centroid location (Appendix B) but not image smear. Comparison of pairs of results using different high-resolution images - which are subject to different smear, Table 3 – does not reveal any systematic differences. We conclude that, as with single images, smear plays a minor role. Differences between the different techniques are also smaller than the estimated uncertainties. As a result, despite the preliminary nature of our analysis we regard the estimated errors in Table 6 as likely to be realistic. For Pluto, taking a simple mean and assuming the largest estimated errors as representative, we estimate the spherical radius to be $1188.3 \pm 1.6$ km (2-$\sigma$). For Charon we find a radius of $606.0 \pm 1.0$ km (2-$\sigma$). Relative to the resolution of the best images, the estimated errors are both about 0.4-0.5 pixels.

As a reality check, we can compare this estimated uncertainty with previous radius determinations using comparable numbers of images. For the five Uranian satellites, the estimated radius uncertainty relative to the resolution of the best image is in all cases about 0.5 pixels (Thomas 1988), the same as our estimate. The estimated uncertainty in the mean radius for Europa based on limb profiles is 0.3 km or 0.1 pixels (Nimmo *et al.* 2007), suggesting that our uncertainty estimate is if anything conservative. For Europa a control point network analysis yielded a radius uncertainty of 0.65 km (Davies *et al.* 1998).

Table 6 also lists the maximum possible flattening *f=(a-c)/a*, where *f* is constrained to be positive. For Pluto and Charon the maximum positive values of *f* are +0.6% (except for method C) and +0.5%, respectively, but in both cases a completely spherical body is consistent with the observations.

### 5. Discussion

Pre-*New Horizons* estimates of Pluto's radius yielded large uncertainties, e.g., a range of 1169-1193 km [Lellouch *et al.* 2009] or $1179 \pm 24$ km [Young and Binzel 1994].

Our LORRI-derived value of 1188.3 ± 1.6 km (2-σ) is at the upper end of the previous estimates. It agrees with the occultation-derived values of 1189± 2 km derived from the FSS (Appendix C) and 1191± 3 km from ALICE (Appendix D), and with the REX-derived occultation value of 1189.9±0.4 km [Gladstone *et al.,* 2016]. Pluto is therefore larger than the Kuiper Belt object Eris, which has a radius of 1163 ± 6 km [Sicardy *et al.* 2011].

For Charon, our estimated radius of 606.0 ± 1.0 km (2-σ) may be compared with Earth-based occultation estimates of 606 ± 1.5 km [Person *et al.* 2006] and *R*= 603.6 ± 1.4 km [Sicardy *et al.* 2006]. The three estimates all (just) agree within error. Person *et al.* [2006] also proposed an oblateness of 0.006 ± 0.003, whereas in contrast we are only able to establish an upper bound on oblateness of 0.005.

Person *et al.* [2006] suggested that some discrepancies in the occultation results could be due to local topography on Charon; inspection of *New Horizons* stereo topography lends some weight to this suggestion [Zangari *et al.* 2016]. On the other hand, the values derived for Charon's radius of 619±0.5 km from the FSS (Appendix C) and 619 or 610±7 km from ALICE (Appendix D) are outside the LORRI-derived estimate. These discrepancies could be due to local topography but are more likely the result of errors in the chord locations (which are well off-centre) arising from uncertainties in the relative position between *New Horizons* and Charon. Further work on this issue is required.

Assuming errors add quadratically, the fractional uncertainty in density, $\Delta\rho/\rho=[(\Delta m/m)^2 + (3\Delta R/R)^2]^{1/2}$ where $\Delta m$ and $\Delta R$ are the uncertainties in mass and radius, respectively. The 1-σ fractional uncertainties in mass are given in Brozovic *et al.* [2015], while the 2-σ fractional uncertainties in radius are 0.13% and 0.17% for Pluto and Charon, respectively. The resulting density estimates and 2-σ uncertainties are 1854 ± 11 kg m$^{-3}$ and 1701 ± 33 kg m$^{-3}$. We note that the uncertainties quoted in Stern *et al.* [2015, Table 1] were calculated incorrectly.

Charon's lower density seems to be real and indicates that either its bulk composition differs from that of Pluto, or that it has more porosity. In general one would expect higher porosity for smaller objects, but so far only limited work has been done to quantify this effect [e.g. Brown 2013, Malamud and Prialnik 2015, Bierson *et al.* 2016]. Charon's lower density might alternatively be due to a lower rock fraction, a higher abundance of carbonaceous material [McKinnon *et al.* 1997], or the absence of a present-day subsurface ocean [Hussmann *et al.* 2006, *cf.* Desch *et al.* 2009]. Distinguishing between these possibilities is important, as Charon's bulk composition forms a major constraint on its mode of formation (Section 1).

The predicted present-day flattening values for Pluto and Charon are ~0.05% (Section 1 and see below). Our 2-σ upper bounds on flattening of 0.6% (7km) for Pluto and 0.5% (3km) for Charon place constraints on the thermal/orbital evolution of these bodies [e.g. Robuchon and Nimmo 2011]. Figure 5 shows how the predicted flattening for fluid bodies evolves as a function of their separation. We assume the total angular momentum of the system is conserved and Pluto spins down as the orbital separation increases (Charon is assumed to be synchronous). The flattening of each body can then be obtained assuming that they behave as fluids using the Darwin-Radau relation [Murray and Dermott 2000]:

$$f = \frac{5}{2}\left[\frac{\omega^2 a^3}{Gm}\right] \Big/ \left[\left(\frac{5}{2}\left(1 - \frac{3}{2}\frac{I}{mR^2}\right)\right)^2 + 1\right]$$

where ω is the rotation angular frequency, $a$ is the semi-major axis, $m$ is the mass, $R$ is the radius and $G$ the gravitational constant. The moment of inertia $I$ is obtained assuming a two-layer body with densities of 3400 kgm$^{-3}$ and 950 kgm$^{-3}$ for rock and ice, respectively, with the bulk density constrained to equal the measured value. The corresponding core radii are 853.2 and 408.7 km for Pluto and Charon. Note that this expression ignores the effects of tides and is thus only approximately correct, particularly for Charon. At the present day, Charon is expected to be more tidally-distorted than Pluto, with the predicted quantity *(b-c)/(a-c)* being 0.76 and 0.27 for Pluto and Charon, respectively, assuming hydrostatic equilibrium (McKinnon et al. 2014, and in prep). But all present-day distortions are <1km, too small to be detectable in our analysis.

The absence of detectable flattening at Pluto implies that its interior must have been warm and/or deformable during or subsequent to the bulk of system orbital evolution [Stern *et al.* 2015]. The large stresses associated with spin-down should lead to pronounced global tectonic patterns [Barr and Collins 2015], which are not readily reconciled with the available observations [Moore *et al.,* 2016]. The early evolution of Pluto is thus not recorded in either its shape or on its surface. Unfortunately, the duration of this early epoch depends on the rate of dissipation inside Pluto, which is poorly known but is probably a few Myr [Dobrovolskis *et al.* 1997, Cheng *et al.* 2014, Barr and Collins 2015].

Numerical models show that only in cases when Pluto does not develop a subsurface ocean can a fossil bulge be maintained [Robuchon and Nimmo 2011]. Even in these cases, the extreme stresses associated with spin-down will cause the lithosphere to fracture and thus limit the size of any bulge which can be recorded [McKinnon and Singer 2014]. Thus, although a Pluto with a subsurface ocean is *consistent* with the absence of a fossil bulge, a subsurface ocean is not *required* by this observation.

Charon is comparable in size to Iapetus, a moon of Saturn which shows a very pronounced (~5%) fossil bulge [Castillo-Rogez *et al.* 2007]. The absence of a comparable bulge on Charon is probably due to the fact that Charon reached its present-day spin rate much faster than Iapetus, and was thus too warm and/or deformable to record earlier spin states.

There are various ways in which this work could be followed up. First, there are undoubtedly aspects of our methods and error analysis that could be refined with future work. More broadly, we have focused here primarily on LORRI images. We have paid less attention to other constraints, such as occultations, even though in the case of Charon there is an as yet unresolved discrepancy between the imaging and occultation results. Ultimately, all constraints, including local topography and control points, should be folded into a global solution. Conversely, limb topography, especially for Charon, may be useful for reducing warping in stereo- or photoclinometrically-derived topography. On other satellites, the power spectral behavior of topography, based on limb profiles, has been used to infer effective elastic thicknesses [Nimmo *et al.* 2011] and it would be of considerable interest to do the same for Pluto and Charon. Analysis of the complex structure of the atmospheric hazes [Stern *et al.* 2015] requires the location of the surface to be accurately determined. Perhaps most fundamentally, the accurate density measurements and limits on the degree of flattening presented here should allow the origin, evolution, and present-day structure of Pluto and Charon to be investigated in unprecedented detail.

**Acknowledgements**. We thank the two reviewers for careful reading. We also thank Bill Owen for his careful determination of LORRI's geometrical distortion, and Brian Carcich for his derivation of the SIP distortion coefficients based on Owen's model. HBT thanks Gabe Rogers (APL), Sarah Flanigan (APL), and Michael Kagan (Adcole, Inc) for help with the FSS data. The authors gratefully acknowledge the support of the NASA *New Horizons* project in facilitating this work.

## 6. References

Albrecht, R. *et al*. 1994. High-resolution imaging of the Pluto-Charon system with the Faint Object Camera of the Hubble Space Telescope. *Astrophys. J. Lett*. 435, 75-78.

Anderson, J.D. et al. 2001. Shape, mean radius, gravity field and interior structure of Callisto, *Icarus* 153, 157-161.

Barr, A.C., Collins, G.C., 2015. Tectonic activity on Pluto after the Charon-forming impact, *Icarus* 246, 146-155.

Bierson, C.J., F. Nimmo, W.B. McKinnon, 2016. Testing for a compositional difference between Pluto and Charon. *LPSC* 47, 2176.

Binzel, R.P., Hubbard, W.B., 1997. Mutual events and stellar occultations, in *Pluto and Charon*, S.A. Stern and D.J. Tholen, eds., Univ. Ariz. Press, pp. 85-102.

Brown, M.E., 2013. The density of mid-sized Kuiper Belt object 2002 UX25 and the formation of the dwarf planets, *Astrophys. J. Lett.* 778, L34.

Brozovic, M. *et al.,* 2015. The orbits and masses of satellites of Pluto, *Icarus* 246, 317-329.

Buie, M.W. *et al.*, 2010. Pluto and Charon with the Hubble Space Telescope II. Resolving changes on Pluto's surface and a map for Charon, *Astron. J.* 139, 1128-1143.

Buie, M.W. *et al.*, 1997. Separate lightcurves of Pluto and Charon, *Icarus* 125, 233-244.

Buie, M.W. *et al.,* 1992. Albedo maps of Pluto and Charon – initial mutual event results, *Icarus* 97, 211-227.

Buratti, B. *et al.* 1995. Modeling Pluto-Charon mutual events. II. CCD Observations with the 60 in. telescope at Palomar Mountain. *Astron. J.* 110, 1405-1419.

Canup, R.M., 2005. A giant impact origin of Pluto-Charon, *Science* 307, 546-550.

Castillo-Rogez, J.C. et al., 2007. Iapetus' geophysics: Rotation rate, shape and equatorial ridge, *Icarus* 190, 179-202.

Cheng, A.F. et al., 2008. Long-range reconnaissance imager on New Horizons, *Space Sci. Rev.* 140, 189-215.

Cheng W.H. et al., 2014. Complete tidal evolution of Pluto-Charon, *Icarus* 233, 242-258.

Davies, M.E. et al., 1998. The control networks of the Galilean satellites and implications for global shape, *Icarus* 135, 372-376.

Dermott, S.F., Thomas, P.C., 1988. The shape and internal structure of Mimas, *Icarus* 73, 25-65, 1988.

Desch, S.J. 2015. Density of Charon formed from a disk generated by the impact of partially differentiated bodies, *Icarus* 246, 37-47.

Desch, S.J. et al. 2009. Thermal evolution of Kuiper belt objects, with implications for cryovolcanism, *Icarus* 202, 694-714.

Dobrovolskis, A.R. et al., 1997. Dynamics of the Pluto-Charon binary, in *Pluto and Charon*, S.A. Stern and D.J. Tholen, eds., Univ. Ariz. Press, pp. 159-192.

Dunbar, R.S. and E.F. Tedesco, 1986. Modeling Pluto-Charon mutual eclipse events 1. First order models. *Astron. J.* 92, 1201-1209.

Elliot, J.L. and L.A. Young, 1991. Limits on the radius and a possible atmosphere of Charon from its 1980 stellar occultation. *Icarus* 89, 244-254.

Elliot, J.L. and L.A. Young, 1992. Analysis of stellar occultation data for planetary atmospheres. 1. Model fitting, with application to Pluto. *Astron. J.* 103, 991-1015.

Eshelman, V.R. 1989. Pluto's atmosphere: Models based on refraction, inversion and vapor-pressure equilibrium. *Icarus* 80, 439-443.

FitzGibbon, A. et al., 1999. Direct least square fitting of ellipses, *IEEE Trans. Pattern Analysis Machine Intelligence* 21, 476-480.

Garrick-Bethell, I. et al., 2014. The tidal-rotational shape of the Moon and evidence for polar wander, *Nature* 512, 181-184.

Gladstone, G.R. et al., 2016. The atmosphere of Pluto as observed by New Horizons, *Science* 351, 1280.

Grundy, W.M. et al. 2007. New Horizons mapping of Europa and Ganymede, *Science* 318, 234-237.

Hussmann, H. et al., 2006. Subsurface oceans and deep interiors of medium-sized outer planet satellites and large trans-neptunian objects, *Icarus* 185, 258-273.

Jackman, C. et al., 2016. New Horizons Optical Navigation on Approach to Pluto, AAS paper 16-083, *39th AAS Guidance, Navigation and Control Conference*, Breckenridge, CO.

Kagan, M., 2003. System description/operation manual: Adcole Fine Sun Sensor Assembly, Pluto-Kuiper Belt Mission.

Lauer, T.R., 1999. Combining undersampled dithered images, *Pub. Astron. Soc. Pacific* 111, 227-237.

Lellouch, E. et al., 2009. Pluto's lower atmosphere structure and methane abundance from high-resolution spectroscopy and stellar occultations. *Astron. Astrophys.* 459, L17-L21.

Malamud, U., Prialnik, D., 2015. Modeling Kuiper belt objects Charon, Orcus and Salacia by means of a new equation of state for porous icy bodies, *Icarus* 246, 21-36.

McKinnon, W.B., Singer, K.N. 2014. The internal structures of Pluto and Charon, and can New Horizons tell? *DPS* 46, 419.07.

McKinnon, W.B. et al., 2014. Cosmogonic constraints from densities in the Pluto system and rotational and tidal figures of equilibrium for Pluto and Charon, *Asteroids Comets Meteors Conf.*, Helsinki, Finland.

McKinnon, W.B., et al., 1997. Composition, internal structure and thermal evolution of Pluto and Charon, in *Pluto and Charon*, S.A. Stern and D.J. Tholen, eds., Univ. Ariz. Press, pp. 270-295.

Millis, R.L. *et al.* 1993. Pluto's radius and atmosphere: Results from the entire 9 June 1988 occultation data set. *Icarus* 105, 282-297.

Moore, J.M. et al., 2016, The geology of Pluto and Charon through the eyes of New Horizons, *Science* 351, 1284-1293.

Murray, C.D., Dermott, S.F., 2000. *Solar system dynamics,* Cambridge Univ Press, 608pp.

Nesvorny, D. et al., 2010. Formation of Kuiper Belt binaries by gravitational collapse, *Astron. J.* 140, 785-793.

Nimmo, F. et al., 2007. The global shape of Europa: Constraints on lateral shell thickness variations, *Icarus* 191, 183-192.

Nimmo, F. et al. 2010. Geophysical implications of the long-wavelength topography of Rhea, *J. Geophys. Res.* 115, E10008.

Nimmo, F. et al., 2011. Geophysical implications of the long-wavelength topography of the Saturnian satellites, *J. Geophys. Res.* 116, E11001.

Noble, M.W. et al., 2009. In-flight performance of the Long Range Reconnaissance Imager (LORRI) on the New Horizons mission, *Proc. SPIE* 7441, 74410Y.

Oberst, J., Schuster, P., 2004. Vertical control point network and global shape of Io, *J. Geophys. Res.* 109, E04003.

Owen, W.M., O'Connell, D., 2011. New Horizons LORRI geometric calibration of August 2006, JPL Interoffice memorandum 343L-11-002.

Pelletier, F. J., et al., 2016. New Horizons Orbit Determination Performance During Approach and Flyby of the Pluto System, AAS paper 16-419, *26th AAS/AIAA Space Flight Mechanics Meeting*, Napa, CA.

Person, M.J. et al., 2006. Charon's radius and density from the combined data sets of the 2005 July 11 occultation, *Astron. J.* 132, 1575-1580.

Person, M.J. *et al.* 2013. The 2011 June 23 stellar occulation by Pluto: airborne and ground observations, *Astron. J.* 146, 83.

Reinsch, K. and M.W. Pakull, 1987. Physical parameters of the Pluto-Charon system. *Astron. Astrophys.* 177, L43-L46.

Reinsch, K. *et al.* 1994. Albedo maps of Pluto and improved physical parameters of the Pluto-Charon system. *Icarus* 108, 209-218.

Robuchon, G., F. Nimmo, 2011. Thermal evolution of Pluto and implications for surface tectonics and a subsurface ocean, *Icarus* 216, 426-439.

Shupe, D.L. et al., 2005. The SIP convention for representing distortion in FITS image headers, Astron. *Data Analysis Software and Systems* 14, vol. 30, P.L. Shopbell, M.C. Britton and R. Ebert, eds.

Sicardy, B. et al. 2006. Charon's size and an upper limit on its atmosphere from a stellar occultation, *Nature* 439, 52-54.

Sicardy, B. et al. 2011. A Pluto-like radius and a high albedo for the dwarf planet Eris from an occultation, *Nature* 478, 493-496.

Slater, D.C. et al., 2005. Radiometric performance results of the *New Horizons' ALICE* UV imaging spectrograph. Proc. SPIE 5906, 590619-1-12.

Steffl, A.J. et al., New Horizons Spice kernels V2.0.1, NH-J/P/SS-SPICE-6-V1.0, NASA Planetary Data System, 2016.

Stern, S.A. et al., 2015. The Pluto system: initial results from its exploration by New Horizons, *Science* 350, 292.

Stern, S.A. et al., 2008. ALICE: The ultraviolet imaging spectrograph aboard the New Horizons Pluto-Kuiper Belt mission. *Space Sci. Rev.* 140, 155-187, 2008.

Tholen, D.J. 2014. The size of Pluto, *DPS* 46, 404.01.

Tholen, D.J. and M.W. Buie, 1988. Further analysis of Pluto-Charon mutual event observations – 1988. *Bull Amer. Astron. Soc.* 20, 807.

Tholen, D.J. and M.W. Buie, 1990. Further analysis of Pluto-Charon mutual event observations – 1990. *Bull Amer. Astron. Soc.* 22, 1129.

Thomas, P.C., 1988. Radii, shapes and topography of the satellites of Uranus from limb coordinates, *Icarus* 73, 427-441.

Thomas, P.C. et al., 1998. The shape of Io from Galileo limb measurements, *Icarus* 135, 175-180.

Thomas, P.C. et al., 2007. Shapes of the Saturnian icy satellites and their significance, *Icarus* 190, 573-584.

Thomason, C.J., F. Nimmo, 2015. Determination of Pluto's radius during the New Horizons encounter, *LPSC* 46, 1462.

Young, E.F. 1992. An albedo map and frost model of Pluto. Ph.D. thesis, Massachusetts Inst. of Technology.

Young, E.F., R.P. Binzel, 1994. A new determination of radii and limb parameters for Pluto and Charon from mutual event lightcurves, *Icarus* 108, 219-224.

Zangari, A.M. et al., 2016. Have stellar occultations probed Charon's chasmata? *LPSC* 47, 1535.

| Reference | Pluto radius (km) | Charon radius (km) | Notes |
|---|---|---|---|
| Dunbar and Tedesco (1986) | 1150±50 | 750±50 | |
| Reinsch and Pakull (1987) | 1100±70 | 580±50 | |
| Tholen and Buie (1988) | 1142±9 | 596±17 | |
| Eshleman (1989) | 1180±23 | - | |
| Tholen and Buie (1990) | 1151±6 | 593±13 | |
| Elliot and Young (1991) | - | >601.5 | |
| Elliot and Young (1992) | 1206±11 | - | Positive thermal gradient |
| Young (1992) | 1191±20 | 642±11 | |
| Buie et al. (1992) | 1150±7 | 593±10 | |
| Millis et al. (1993) | 1195±5 | - | Positive thermal gradient |
| Reinsch et al. (1994) | 1152±7 | 595±5 | Recalibrated semi-major axis |
| Young and Binzel (1994) | 1179.5±23.5 | 629±21 | Recalibrated semi-major axis |
| Albrecht et al. (1994) | 1160±12 | 635±13 | |
| Buratti et al. (1995) | 1155±20 | 612±30 | |
| Person et al. (2006) | - | 606.0±1.5 | |
| Sicardy et al. (2006) | - | 603.6±1.4 | |
| Lellouch et al. (2009) | 1169-1193 | - | |
| Stern et al. (2015) | 1187±4 | 606±3 | *New Horizons* |
| This work | 1188.3±1.6 | 606.0±1.0 | *New Horizons* |

**Table 1.** Previous estimates of Pluto and Charon radii, modified from McKinnon et al. (1997).

| Image name | Image MET | Sub S/C lat $\phi_0$ ° | Sub S/C lon $\theta_0$ ° | Subsol. lat $\phi_s$ ° | Subsol. lon $\theta_s$ ° | Rotation angle $\varphi$ ° | Res. (km/pix) | Alt. name |
|---|---|---|---|---|---|---|---|---|
| Visit70p | 298893474* | 43.01 | 333.84 | 51.54 | 315.09 | 319.30 | 9.8230* | |
| Visit71p | 298959290* | 42.98 | 291.00 | 51.55 | 272.15 | 319.25 | 7.5694* | |
| Visit72p | 298996664* | 42.95 | 266.69 | 51.55 | 247.76 | 229.17 | 6.2893* | |
| Visit73p | 299124574 | 42.54 | 184.01 | 51.55 | 164.31 | 219.72 | 3.8206 | PELR_LORRI OFF_038 |
| Visit74p | 299123689 | 42.55 | 184.57 | 51.55 | 164.88 | 172.77 | 3.8812 | |
| Visit75p | 299127017 | 42.51 | 182.46 | 51.55 | 162.71 | 201.47 | 3.6535 | P_LORRI_FUL LFRAME_1 |
| Visit70c | 298893754* | 43.19 | 153.47 | 51.54 | 134.91 | 319.51 | 9.7820* | |
| Visit71c | 298959599* | 43.06 | 110.32 | 51.55 | 91.94 | 319.55 | 7.5472* | |
| Visit72c | 298996974* | 42.93 | 85.89 | 51.55 | 67.56 | 229.61 | 6.2820* | |
| Visit74c | 299147641 | 40.50 | 350.33 | 51.55 | 329.26 | 315.50 | 2.3148 | |
| Visit76c | 299147776 | 40.49 | 350.26 | 51.55 | 329.17 | 315.47 | 2.3055 | C_LORRI_FU LLFRAME_1 |

**Table 2.** Images used in analysis. MET refers to spacecraft clock time. Here suffix p denotes Pluto and c denotes Charon. Asterisks denote summed images derived by Nyquist sampling of multiple images closely spaced in time (Section 2.2) and with slightly different MET identifiers. The pixel sampling of these images is twice as fine as that of the LORRI source images. Resolutions are determined from spacecraft range assuming an angular resolution of 4.963571 μrad/pixel. The rotation angle is the clockwise angle between the assumed body rotation pole and the image y-axis.

| | $\Delta x$ (pixel) | $\Delta y$ (pixel) | $\Delta$ (pixel) | REQ_ID & Image |
|---|---|---|---|---|
| LOR_0298893474 | -0.48 | -0.22 | 0.53 | PELR_NAV_C4_L1_CRIT_35_02; 70p |
| LOR_0298893504 | 0.05 | -0.33 | 0.34 | PELR_NAV_C4_L1_CRIT_35_02; 70p |
| LOR_0298959290 | -0.53 | -0.02 | 0.53 | PELR_NAV_C4_L1_CRIT_36_01 ; 71p |
| LOR_0298959320 | -0.04 | 0.08 | 0.09 | PELR_NAV_C4_L1_CRIT_36_01; 71p |
| LOR_0298959350 | 0.53 | 0.16 | 0.55 | PELR_NAV_C4_L1_CRIT_36_01; 71p |
| LOR_0298996664 | 0.51 | -0.18 | 0.54 | PELR_NAV_C4_L1_CRIT_37_01; 72p |
| LOR_0298996694 | -0.35 | 0.04 | 0.35 | PELR_NAV_C4_L1_CRIT_37_01; 72p |
| LOR_0298996724 | 0.25 | -0.07 | 0.26 | PELR_NAV_C4_L1_CRIT_37_01; 72p |
| LOR_0298893754 | -0.48 | 0.37 | 0.61 | PELR_NAV_C4_L1_CRIT_35_03; 70c |
| LOR_0298959599 | 0.60 | 0.27 | 0.66 | PELR_NAV_C4_L1_CRIT_36_02; 71c |
| LOR_0298959629 | 0.04 | -0.47 | 0.48 | PELR_NAV_C4_L1_CRIT_36_02; 71c |
| LOR_0298996974 | 0.39 | 0.42 | 0.57 | PELR_NAV_C4_L1_CRIT_37_02; 72c |
| LOR_0298997004 | -0.05 | -0.19 | 0.20 | PELR_NAV_C4_L1_CRIT_37_02; 72c |
| LOR_0299124574 | 0.17 | -0.18 | 0.24 | PELR_LORRIOFF_038; 73p |
| LOR_0299127017 | 0.48 | -0.16 | 0.50 | PELR_P_LORRI_FULLFRAME_1; 75p |
| LOR_0299147641 | 0.87 | 0.21 | 0.90 | PELR_C_LORRI_FULLFRAME_1; 74c |
| LOR_0299147776 | -0.28 | 0.27 | 0.39 | PELR_C_LORRI_FULLFRAME_1; 76p |
| LOR_0299123689 | 0.52 | 0.42 | 0.67 | PELR_PC_AIRGLOW_FILL_2_08; 74p |

**Table 3.** Image smear $\Delta x$ and $\Delta y$ in X and Y direction during individual exposures. Note that some images are produced by combining several exposures (Section 2.2); REQ_ID is an identifier to aid in image retrieval from the PDS archive.

|  | Method | N | R (pix) | R (km) | $x_0$ (pix) | $y_0$ (pix) | RMS (km) |
|---|---|---|---|---|---|---|---|
| Visit 70p 9.8230* km/pix | A 30% | 171 | 121.2±0.2 | 1190.5±2.1 | 269.4 | 266.1 | 3.17 |
|  | B 50% | 340 | 121.6±0.2 | 1194.5±2.3 | 271.0 | [777.1] | 2.79 |
|  | C picks | 221 | 121.7±1.1 | 1195.5±11 | 269.6 | [778.4] | 13.3 |
|  | C unwrap | 480 | 120.7±0.8 | 1185.6±7.9 | 269.2 | [777.0] | n/a |
| Visit71p 7.5694* km/pix | A 30% | 416 | 156.9±0.3 | 1187.6±2.4 | 277.3 | 264.4 | 3.70 |
|  | B 50% | 466 | 156.5±0.3 | 1184.6±2.1 | 278.1 | [775.0] | 2.77 |
|  | C picks | 569 | 156.8±1.0 | 1186.0±7.9 | 276.6 | [776.2] | 10.6 |
|  | C unwrap | 480 | 156.8±1.0 | 1186.8±7.6 | 277.1 | [775.8] | n/a |
| Visit72p 6.2893* km/pix | A 30% | 326 | 188.9±0.4 | 1188.0±2.4 | 258.6 | 231.6 | 3.02 |
|  | A 40% | 487 | 188.7±0.3 | 1186.8±2.0 | 258.9 | 230.9 | 2.39 |
|  | B 50% | 561 | 189.4±0.3 | 1191.2±1.9 | 258.8 | [740.9] | 2.83 |
|  | C picks | 554 | 187.2±1.3 | 1177.4±8.1 | 255.7 | [744.7] | 9.87 |
|  | C unwrap | 420 | 188.7±0.3 | 1186.7±1.9 | 255.2 | [743.8] | n/a |
| Visit74p 3.8812 | C unwrap | 540 | 307.5±0.9 | 1193.5±3.5 | 508.0 | 522.1 | n/a |
| Visit73p 3.8206 km/pix | A 30% | 843 | 311.4±0.3 | 1189.7±1.1 | 514.2 | 548.2 | 1.33 |
|  | A, dis, 30% | 865 | 311.3±0.3 | **1189.4±1.0** | 514.2 | 548.1 | 1.31 |
|  | A 40% | 790 | 311.1±0.3 | **1188.6±1.3** | 514.2 | 548.2 | 1.32 |
|  | B 50% | 438 | 311.7±0.2 | **1190.9±0.7** | 514.3 | 546.4 | 1.04 |
| Visit75p 3.6535 km/pix | A 30% | 758 | 325.4±0.4 | 1188.8±1.4 | 439.5 | 514.7 | 1.54 |
|  | A, dis, 30% | 694 | 325.2±0.4 | **1188.1±1.4** | 439.7 | 514.7 | 1.56 |
|  | B, dis, 50% | 677 | 325.1±0.3 | **1187.8±1.2** | 439.8 | 513.2 | 1.07 |
|  | A 40% | 599 | 326.0±0.5 | **1191.0±1.9** | 439.7 | 513.5 | 1.17 |
|  | C picks | 677 | 325.6±1.1 | **1189.6±4.1** | 437.8 | 514.5 | 4.39 |
|  | C unwrap | 420 | 325.0±0.3 | 1187.2±1.1 | [585.4] | 515.9 | n/a |

**Table 4**. Best-fit Pluto radii $R$ from invididual images. Methods (A,B,C) are described in Section 3.1; for method C, two different approaches were employed. For A,B percentages refer to threshold value $f_{thresh}$ for identifying limb, and 'dis' refers to images where a correction for camera distortion has been applied (Section 3.3). Except for the `C unwrap' method uncertainties (2-σ) in $R$ were set by an RMS misfit χ =1.044 $χ_{min}$ (see Section 3.3). The inferred centre of the body is $x_0,y_0$; values in square brackets reflect different image pixel conventions used. $N$ is the total number of limb points used in the fit. Radii in bold were used to obtain a naïve estimate of the actual radius (see text).

|  | Method | N | R (pixel) | R (km) | $x_0$ (pix) | $y_0$ (pix) | RMS (km) |
|---|---|---|---|---|---|---|---|
| Visit70c 9.7820* | A | 147 | 61.5±0.3 | 601.6±2.5 | 258.4 | 254.9 | 3.72 |
|  | B | 158 | 62.3±0.1 | 609.4±1.0 | 259.6 | [766.3] | 2.01 |
| Visit71c 7.5472* | A | 237 | 79.8±0.3 | 602.3±2.2 | 259.3 | 253.0 | 2.91 |
|  | B | 260 | 80.4±0.1 | 606.8±1.0 | 260.2 | [764.4] | 1.29 |
| Visit72c 6.2820* | A | 233 | 96.0±0.3 | 603.1±2.0 | 120.5 | 123.95 | 2.75 |
|  | B | 123 | 96.2±0.2 | 604.3±0.8 | [345.0] | [503.8] | 1.21 |
| Visit76c 2.3055 km/pixel | A | 704 | 262.7±0.5 | **605.7±1.3** | 535.75 | 494.1 | 1.73 |
|  | B | 522 | 262.9±0.6 | **606.1±1.3** | 536.4 | 493.3 | 1.76 |
|  | C, picks | 1050 | 263.3±0.7 | **607.0±1.7** | 535.0 | 494.7 | 2.46 |
|  | C, unwrap | 539 | 262.7±0.6 | 605.7±1.4 | 536.5 | 493.4 | n/a |
| Visit 74c 2.3148 km/pixel | A | 868 | 261.6±0.6 | **605.6±1.3** | 484.8 | 485.4 | 1.80 |
|  | B | 705 | 262.0±0.5 | **606.5±1.1** | 485.65 | 484.5 | 1.49 |
|  | C, picks | 968 | 262.2±0.8 | **607.0±1.9** | 484.05 | 486.05 | 2.70 |
|  | C, unwrap | 539 | 261.4±0.3 | 605.1±0.7 | [498.5] | [526.3] | n/a |

**Table 5**. As for Table 4, but for Charon radii. Methods A and B used a 50% threshold throughout.

| **Spherical fits** | | $R$ (km) | | RMS (km) | $N$ |
|---|---|---|---|---|---|
| Pluto: | | | | | |
| 70p-72p & 73p,dis, A 30% | | 1189.0 ± 1.0 | | 2.72 | 1778 |
| 70p-72p & 75p,dis, A 30% | | 1188.0 ± 0.9 | | 2.82 | 1612 |
| 70p-72p & 75p,dis, A 40% | | 1187.6 ± 1.6 | | 4.94 | 1792 |
| 70p-72p & 75p,dis, B 50% | | 1188.4 ± 1.2 | | 4.10 | 2044 |
| 70p-72p,75p, C picks | | 1188.4 ±1.6 | | 4.83 | 1602 |
| **Recommended value** | | **1188.3 ±1.6** | | | |
| | | | | | |
| Charon: | | | | | |
| 70c-72c & 74c, A 50% | | 605.5 ± 1.0 | | 2.94 | 1475 |
| 70c-72c ,74c, B 50% | | 606.5 ± 1.0 | | 2.01 | 1246 |
| **Recommended value** | | **606.0 ± 1.0** | | | |
| | | | | | |
| **Oblate fits** | $a$ (km) | $c$ (km) | $(a-c)/a$ max. | RMS (km) | $N$ |
| Pluto: | | | | | |
| 70p-72p & 73p,dis, A 30% | 1189.0 ± 1.6 | 1189.2 ± 3.4 | 0.4% | 2.72 | 1778 |
| 70p-72p & 75p,dis, A 30% | 1188.2 ± 1.8 | 1187.8 ± 3.8 | 0.5% | 2.82 | 1612 |
| 70p-72p & 75p,dis, A 40% | 1186.8 ± 3.4 | 1189.2 ± 5.0 | 0.6% | 4.91 | 1792 |
| 70p-72p & 75p,dis, B 50% | 1188.2 ± 2.5 | 1188.8± 4.5 | 0.6% | 4.05 | 2044 |
| 70p-72p,75p, C picks | 1188.2±3.0 | 1189.0 ±4.0 | 0.7% | 4.83 | 1602 |
| | | | | | |
| Charon: | | | | | |
| 70c-72c & 74c, A 50% | 604.5 ± 1.5 | 607.0 ± 3.5 | 0.4% | 2.90 | 1475 |
| 70c-72c & 76c, A 50% | 605.0 ± 1.5 | 606.5 ± 3.5 | 0.5% | 2.97 | 1321 |
| 70c-72c & 74c, B 50% | 606.5 ± 1.0 | 607.0 ± 2.5 | 0.4% | 1.98 | 1246 |

**Table 6**. Fits to combined projected images (Section 4). Left-hand column gives images used and method. Here percentages refer to the threshold value adopted and 'dis' indicates that the image was corrected for camera distortion (Section 3.3). Both spherical and oblate shapes are assumed. The quantity *(a-c)/a* max. is the maximum positive flattening. Uncertainties (2-σ) were set by an RMS misfit $\chi=1.1\chi_{min}$ (see Section 3.3) with images weighted according to their resolution. Best-fit axis values were determined in increments of 0.2 km. *N* is the number of points used in the fit. For the C picks only, individual points with radial distances outside +/- 10 km of the nominal radius were excluded from the analysis (this also reduces the total *N* relative to Table 4).

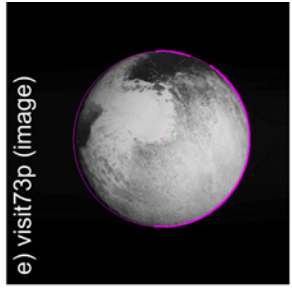
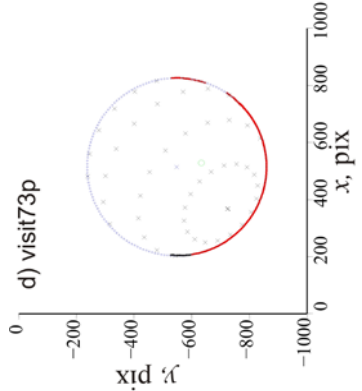
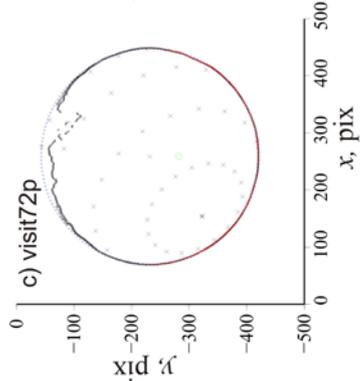
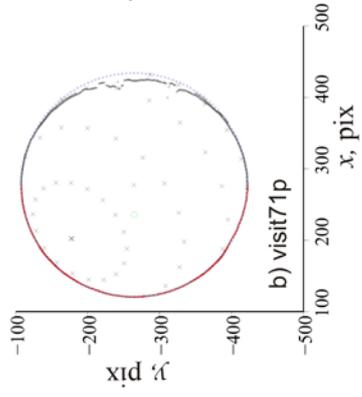
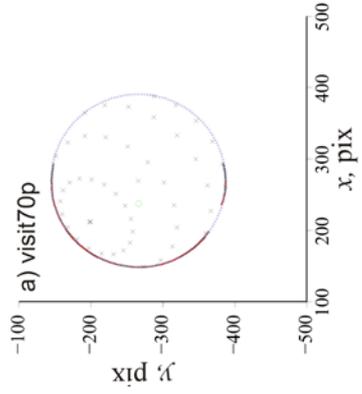
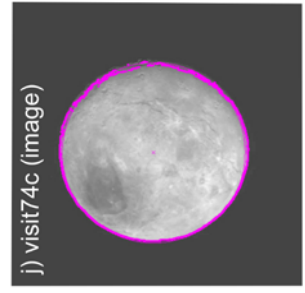
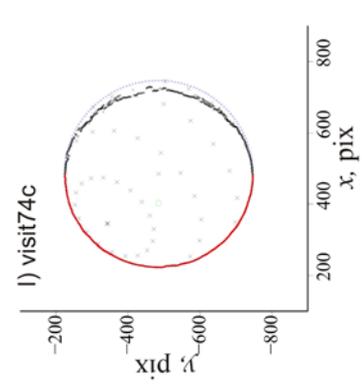
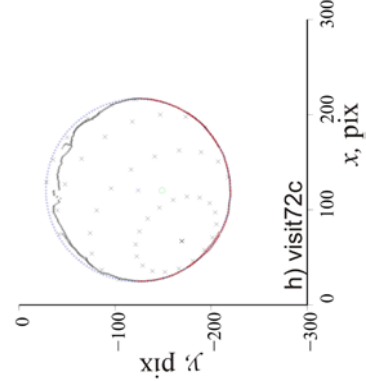
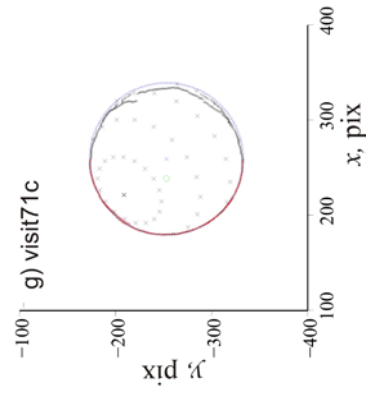
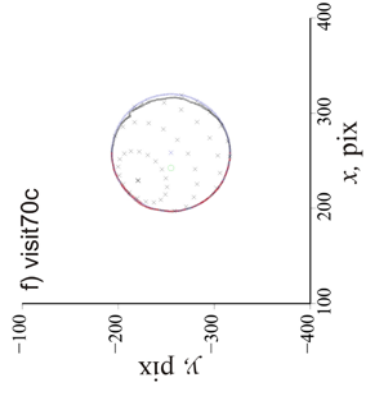

**Figure 1.** Edge picks obtained using method A and 30% threshold from individual images (Table 2). Coordinate system is that of original LORRI image. Black and red points indicate individual picks; red points are calculated to be on the illuminated limb. Note that in a) and d) some points were manually deleted because the edge-picking algorithm sometimes fails when large albedo contrasts are present. Blue cross and green circle denote sub-spacecraft and sub-solar point, respectively; black geographic grid-points at 30° intervals are also shown. Blue points plot best circular fit to illuminated limb (Table 4). Panels e) and i) show raw images with initial picks performed by method A.

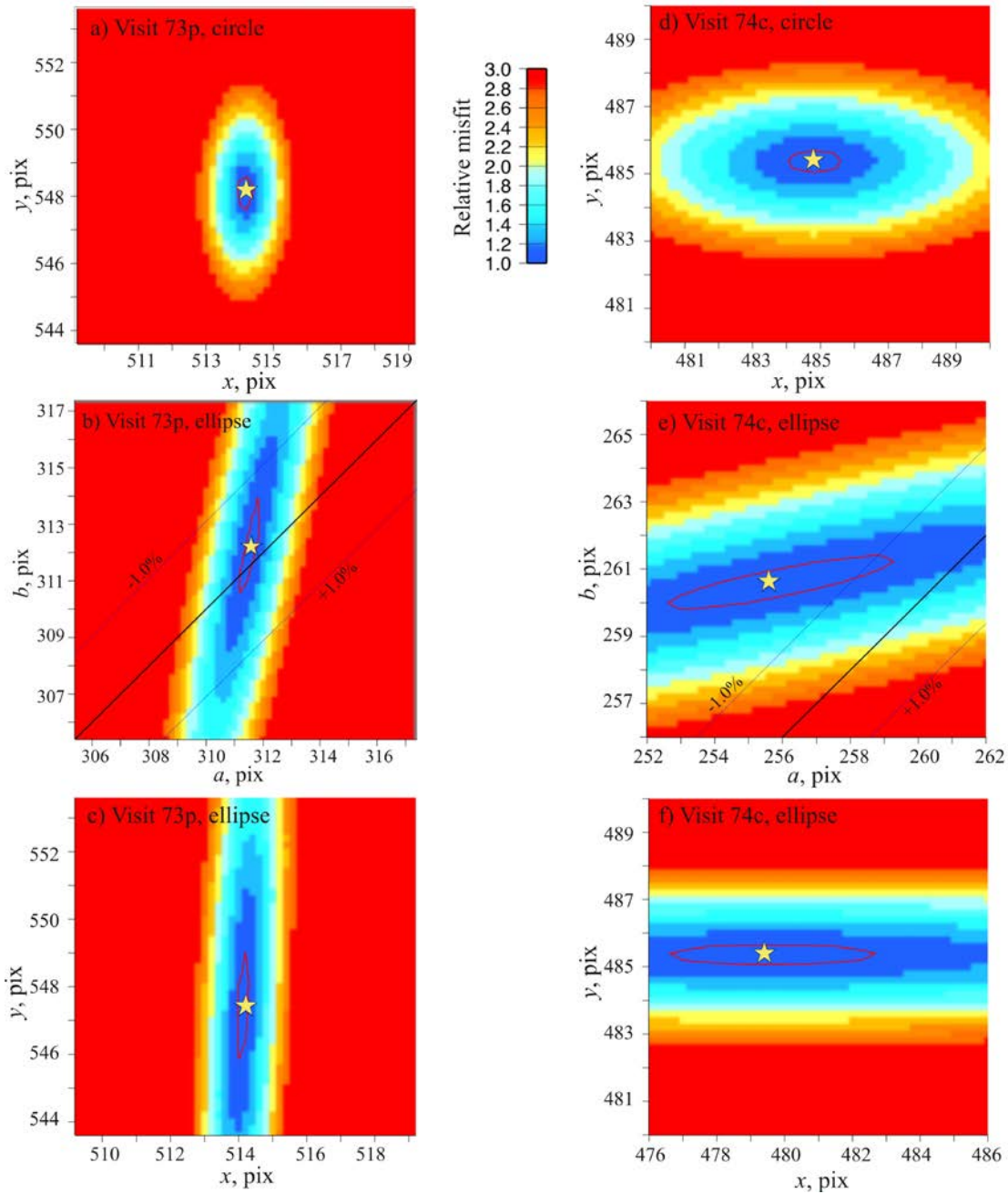

**Figure 2.** a) Relative misfit ($\chi/\chi_{min}$; see text) as a function of assumed circle centre ($x_0,y_0$) for visit 73p (Pluto; Fig 1d). For each point, the local best-fit radius was used. The red contour indicates the 2σ error ellipse (see text); the star indicates the best-fit value (Table 3). Note the elongation of the error ellipse in the *y* direction. b) Relative misfit of assumed best-fit ellipse axes (*a,b*). For each point,

the local best-fit $(x_0,y_0)$ values were used. Note that $a$ is better constrained than $b$. Inclined lines indicate flattening of 0, +1% and -1%. c) Relative misfit of assumed best-fit ellipse centre $(x_0,y_0)$. For each point, the local best-fit $(a,b)$ axis values were used. Note that $x_0$ is better constrained than $y_0$. d)-f) As for a)-c), but for 74c (Charon). In this case, $y_0$ and $b$ are better constrained than $x_0$ and $a$.

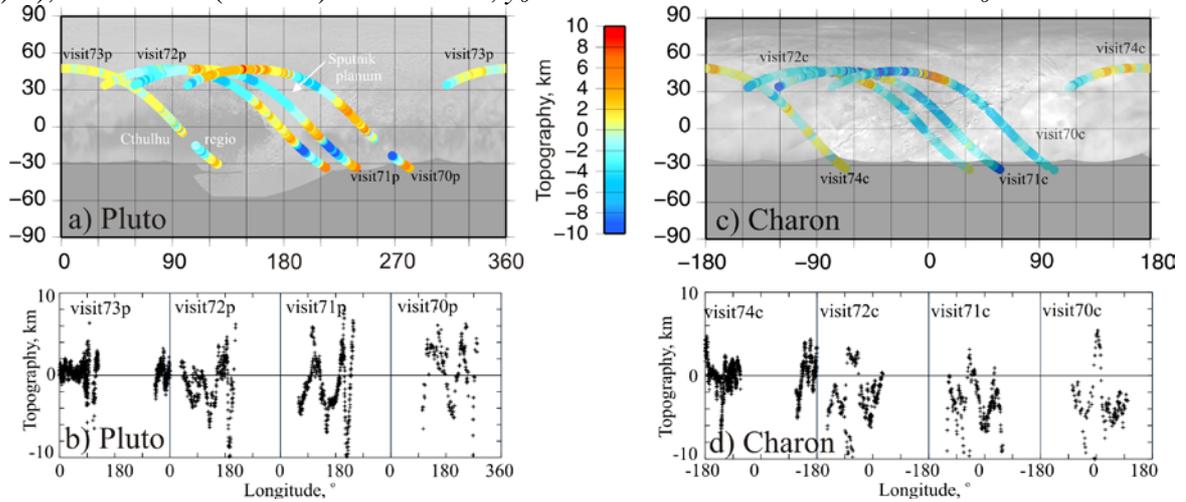

**Figure 3.** a) Pluto limb picks (Figs 1a-d) projected onto surface, simple cylindrical projection. The topography is relative to an oblate spheroid with $a=b=1189.0$ and $c=1189.2$ km (Table 6). Background image is mosaic modified from Moore *et al.* (2016) b) Topography as a function of longitude for the picks shown in a). c) Charon limb picks (Figs 1f-i) projected onto surface. The topography is relative to a sphere of radius 605.5 km (Table 6). d) As for b).

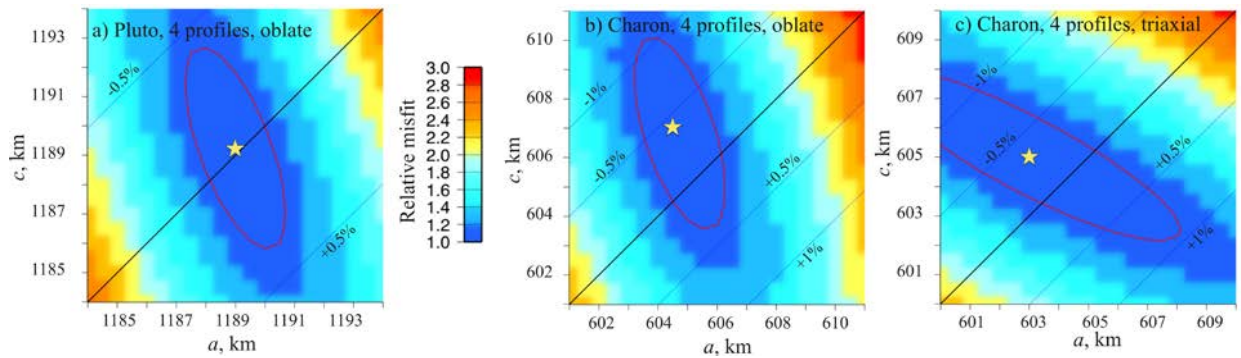

**Figure 4.** a) Relative misfit ($\chi/\chi_{min}$) as a function oblate spheroid axes $(a,c)$ for the projected limb picks shown in Fig 3a. Red line indicates $2\sigma$ error ellipse ($1.1\chi_{min}$), yellow star is best-fit value. Slanted lines show +0.5%,0,-0.5% flattening. b) As for a), but for Charon (Fig 3c). Slanted lines show +1%,+0.5%,0,-0.5%,-1% flattening. c) As for b), but with Charon fit by a hydrostatic triaxial ellipsoid where $(a-c)=4(b-c)$.

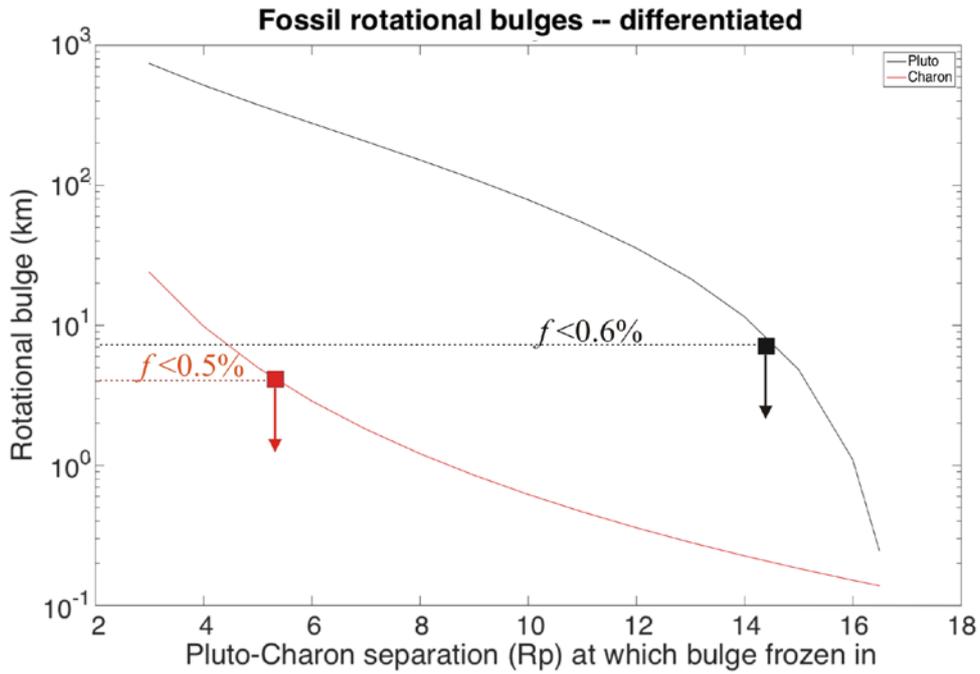

**Figure 5.** Flattening as a function of separation for Pluto and Charon, assuming fluid bodies with differentiated interiors (see text). Horizontal dotted lines denote upper bounds on flattening obtained by analysis of images. The timescale for evolution depends on the poorly-known dissipation inside Pluto, but is probably of order a few million years [Dobrovolskis et al. 1997].

Appendix A: Validation of limb-picking techniques

The possibility of systematic error in the size measurement of Pluto was a strong concern, largely due to the extreme contrast in albedo on its surface. In principle, all of the methods could be biased in their determination by limb darkening on the illuminated disk. We devised a test based on synthetic images to determine the sensitivity of each method to limb-darkening effects and quantify any systematic errors.

The synthetic images were generated based on the Pluto map from Buie *et al.* [2010]. In that work there were six different sets of Hapke photometric parameters that are coupled to the six different maps. Each set (A-F) was used to generate synthetic images. These images were constructed so as to be indistinguishable from the real data except for the use of a lower resolution map.

The synthetic images were calculated taking into account the single-scattering albedo, the other global Hapke parameters ($h$, $P(g)$, $B_0$, $\bar{\theta}$), and the image geometry. A model image was computed at a factor of 10 finer sampling than the LORRI images. After computation, the image was downsampled by pixel-integration to match the scale of the observed image. The model image was computed as bi-directional reflectance which was then converted to I/F, then to the signal detected by the LORRI detector. The final step was to convolve the down-sampled image with the observed PSF of LORRI. The PSF for the center of the array was used for all synthetic images.

Because of the way the images were computed, the radii (in km and pixels) were different for each photometric scheme: $R_A$= 1152.4 km (93.79 pix), $R_B$= 1162.8 km (94.63 pix), $R_C$ = 1167.2 km (94.99 pix), $R_D$ = 1158.3 km (94.26 pix), $R_E$ = 1145.5 km (93.22 pix), and $R_F$ = 1175.1 km (95.63 pix). We emphasize that these are *synthetic* radii, not the estimated true radii.

The results obtained by the three different techniques are displayed in Fig A1 below. Methods A and B show similar behavior: a larger threshold generally results in a systematically smaller estimated radius, as expected (small deviations are due to manual removal of points required for larger thresholds). For method A, thresholds in the range 30-40% perform best, while for method B, the best threshold is 50%. This difference most likely arises because of the differing range of azimuthal angles selected by the two techniques. Method C has no tunable parameters but generally results in radii that are systematically small, by about 0.4% (0.5 pixels) on average.

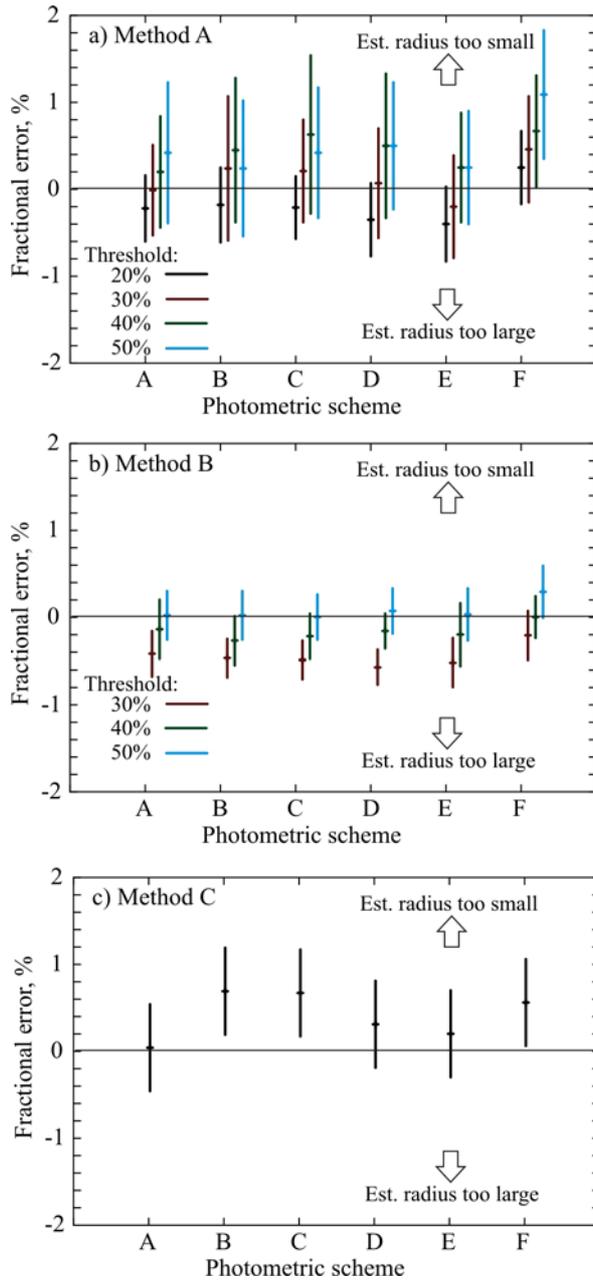

**Fig A1.** Performance of different radius determination methods on synthetic Pluto images. Six different images were created, with different photometric schemes (see text). For methods A and B, a threshold (percentage value) must be specified. For methods A and B the uncertainty was estimated by setting the acceptable RMS misfit $\chi=\chi_{min}(1 + 2\,(2/N)^{1/2})$ with $N$ the number of points.

Appendix B: Error ellipse estimation

In order to assess the uncertainties associated with deriving a shape from a set of limb picks, we generated synthetic limb picks (based on a known "true" shape) and carried out the same fitting procedure as used for the real limb picks. Note that this approach does not investigate uncertainties associated with deriving the actual picks themselves e.g. it does not investigate systematic errors in the images such as those caused by camera distortion or image smear (see Section 3.3).

### B.1 Synthetic limb pick generation

A set of synthetic limb picks contains three components: short wavelength topography; random noise; and the modeled long-wavelength shape of the body.

Synthetic topography was generated on a sphere using the same approach as described in Nimmo *et al.* [2011]. For this particular case, topography was generated between spherical harmonic degrees $l$=20 and $l$=120, with a power law slope of -1.5 and ignoring any flexural deformation. The amplitude of the topography is denoted by the dimensionless parameter $h$

Random noise was added to this topography. In what follows it was taken to be uniformly distributed and to have a peak-to-peak amplitude of 0.4 pixels.

Lastly, the long wavelength shape of the body was added. For a single synthetic limb profile, an ellipse was defined as having axes $a,b$ and central coordinates $x_0,y_0$, and this shape was then added to the synthetic topography + noise. In the case where multiple synthetic profiles were fit simultaneously, the real (lat,long) limb coordinates were projected onto the synthetic topography + noise, and the shape of the oblate spheroid (axes $a,c$) was added. This approach ensures that the differing pixel resolution of different images is taken into account. For the examples shown below we took the long axis to be 313 pixels and the short axis to be 312 pixels (an oblateness of 0.3%).

Figure B1 below shows 3 synthetic profiles compared with an actual one. Fig B1a shows initial limb picks derived from visit73p with a cutoff of 30% (the data gaps are where low-albedo regions prevent picking of the limb – see Fig 1d). Here the picks are plotted relative to a circle of radius 311.1 pixels or 1188.4 km. The RMS deviation from this best-fit shape, $\sigma_{rms}$, is 1.57 km. Figs B1b-d are three synthetic limb picks with different values of topographic amplitude, as represented by the parameter $h$. Visual comparison and the values of $\sigma_{rms}$ suggest that an $h$ in the range 1.0-1.5 is appropriate. Below we will generally take $h$=1.5 as being conservative (rougher than the actual data).

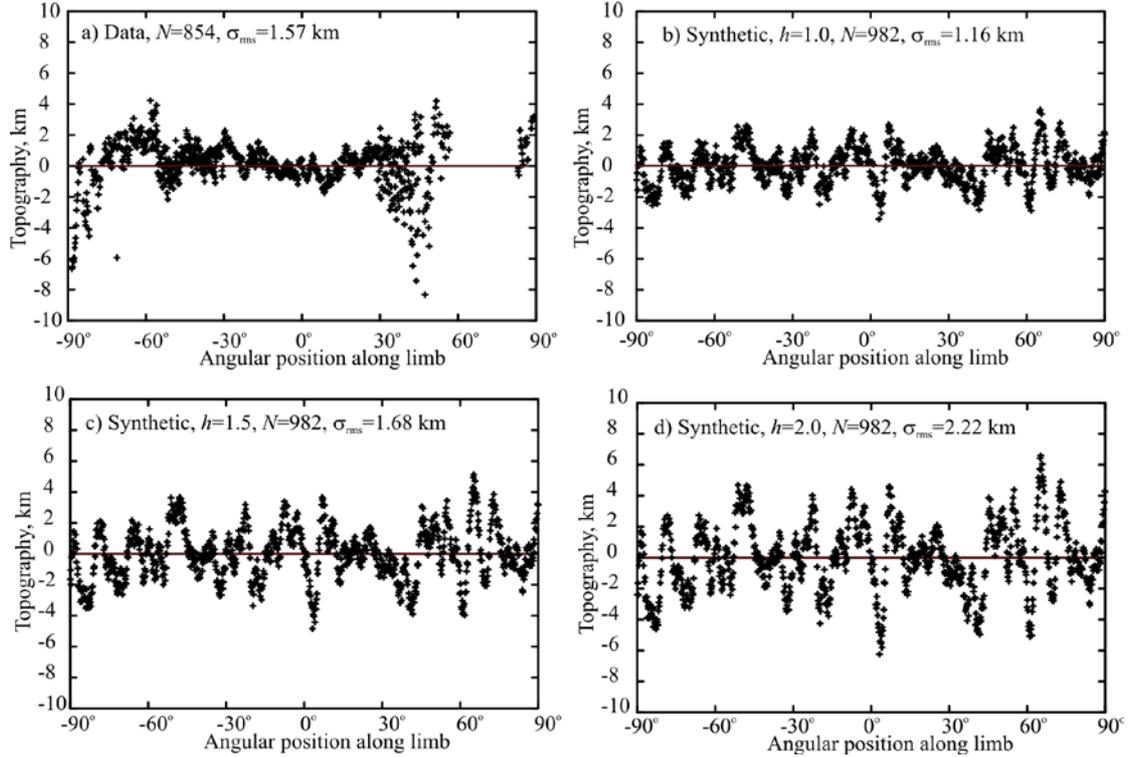

**Figure B1.** a) Initial limb picks from visit73p. Residual topography is plotted relative to a circle of radius 311.1 pixels. Resolution 3.82 km/pixel. $N$ is the number of data points and $\sigma_{rms}$ is the RMS deviation from the best-fit shape. b) –d) Synthetic limb picks, generated as described in the text, with different values of topographic amplitude (denoted by $h$).

### B.2 Fitting single sets of picks

For a single set of synthetic picks, we then fit them to a shape using a grid-search exactly as with the real data (Section 3.2). Doing so generates a best-fit elliptical shape ($a_{min}, b_{min}$), centre coordinates ($x_{0,min}, y_{0,min}$) and RMS misfit $\chi_{min}$ between the observations and the data (equation 2).

Because of the topography and noise, in general the best-fit parameters ($a_{min}, b_{min}$ etc.) are *not* the same as the actual parameters used to generate the synthetic ($a_{true}, b_{true}$ etc.), and the misfit associated with the true parameters $\chi_{true}$ is larger than $\chi_{min}$.

By fitting many different synthetic realizations, we can then ask: how much larger is $\chi_{true}$ than $\chi_{min}$ at the 68% (1-$\sigma$) or 95% (2-$\sigma$) limit? In other words, how large does $\chi$ have to be to ensure that the true parameters fall within the error bounds for 68% or 95% of the realizations? We can then use this particular $\chi$ value as our 1-$\sigma$ or 2-$\sigma$ confidence interval. In practice, this was done by reporting the quantity $\chi_{true}/\chi_{min}$ at the 68% or 95% level.

Similarly, the same realizations can also be used to determine how far $a_{true}, b_{true}$ etc. depart from $a_{min}, b_{min}$ etc. at the 68% or 95% level. Because the realizations produced a distribution of $|\Delta a|=|a_{true}-a_{min}|$ we report the 68% or 95% value of this quantity. These quantities give an estimate of the likely uncertainty in *a,b* etc.

Table B1 summarizes the results for different synthetic topography amplitudes and numbers of data points (*N*). Here a north-south limb profile was assumed, with only the eastern limb illuminated. As expected, rougher topography results in (slightly) larger uncertainties in *a* and *b*. Note, however, that the results are not very sensitive to the number of data points (*N*). This is because the dominant topographic wavelength is long compared with the sampling frequency. For the most likely case (*h*=1.5) the maximum permitted value of $\chi$ at the 1-$\sigma$ level is 1.022 $\chi_{min}$ and the 1-$\sigma$ uncertainties in *a* and *b* are 0.8 and 0.2 pixels (the difference is due to the assumed viewing geometry). We take the corresponding 2-$\sigma$ value to be approximately 1.044 $\chi_{min}$.

| Synthetic | N | $\chi_{true}/\chi_{min}$ (68%) | $|\Delta a|$ (68%)(pix) | $|\Delta b|$ (68%) (pix) |
|---|---|---|---|---|
| *h*=1 | 982 | 1.018 | 0.4 | 0.1 |
| *h*=1.5 | 982 | 1.022 | 0.8 | 0.2 |
| *h*=1.5 | 327 | 1.020 | 0.8 | 0.1 |
| *h*=2.0 | 982 | 1.022 | 1.1 | 0.2 |

**Table B1.** Results of shape-fitting to 30 synthetic realizations of single limb profiles. $|\Delta a|$ and $|\Delta b|$ give the expected deviation of the best-fit *a,b* from their true values (in pixels) at the 1-$\sigma$ level. $\chi_{true}/\chi_{min}$ gives the misfit ratio (the misfit with the true parameters compared to the minimum misfit) which encompasses 68% of realizations.

Figure B2a shows a plot of the misfit $\chi/\chi_{min}$ as *a,b* are varied for the synthetic limb picks shown in Fig B1c. Because the limb profile runs north-south and only the eastern half of the limb is illuminated, the horizontal axis (*a*) is much less well-determined than the vertical axis (*b*). The star denotes the best-fit (*a,b*) solution and the circle denotes the "true" solution. The deviations in *a* and *b* are 0.8 and 0.2 pixels, respectively. The red contours denote the 1-$\sigma$ and approximate 2-$\sigma$ uncertainty ellipses, using the values given in Table B1. Note that the true solution lies within the 1-$\sigma$ ellipse (as it will do for two-thirds of the realizations); note also that the 0.3% "true" oblateness cannot be distinguished from sphericity at the 1-$\sigma$ level. Figure B2b is similar, but plots the misfit for the centre coordinates $x_0, y_0$. As expected, there is much larger uncertainty in $x_0$ than $y_0$, owing to the limb geometry: $x_0$ and *a* trade-off against each other. A single profile cannot determine, within uncertainty, whether the shape is oblate or not.

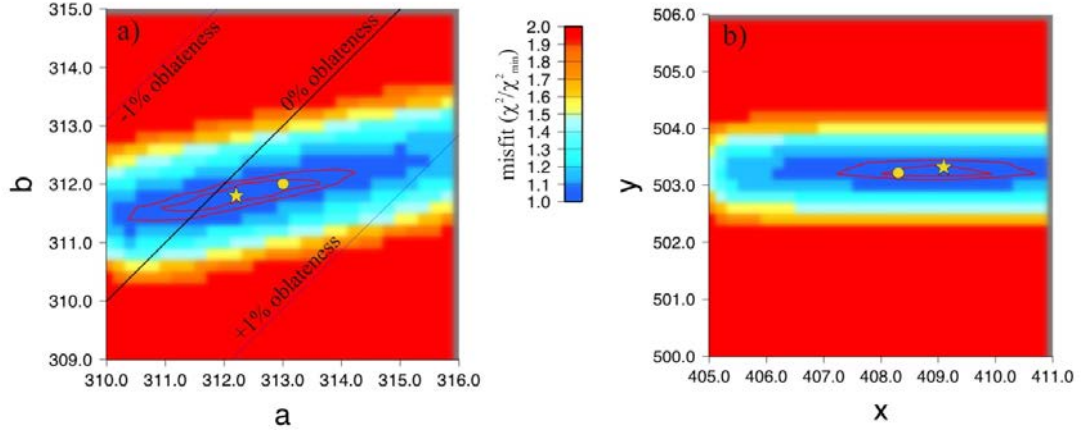

**Figure B2** a). Misfit ($\chi/\chi_{min}$) as a function of *a,b* when fitting the limb picks shown in Fig B1c. For each *a,b* combination the corresponding best-fit centre coordinate $x_0,y_0$ is used to calculate the misfit. The star represents the minimum misfit solution ($\chi_{min}$=0.439 pixels), the circle represents the true synthetic solution ($\chi_{true}$=0.443 pixels). The red contours are given by $\chi/\chi_{min}$=1.022 (1-$\sigma$, Table B1) and 1.044 (which is approximately 2-$\sigma$). b) As for a), but showing how misfit varies with $x_0,y_0$.

### B.3 Fitting multiple projected profiles

A similar analysis can be performed with multiple profiles projected onto a sphere, exactly as with the real data (Section 4). Below we used the projected initial pick locations from visits 70p-72p & 75p but replaced the actual topography with our synthetic topography.

Table B2 summarizes the results based on 30 different realizations. Compared with the single profile, the uncertainties in *a* are greatly reduced, because the multiple profiles stop the tradeoffs between $x_0$ and *a* associated with a single profile (Fig B2). In fact, *c* is more poorly constrained than *a* because the limb profiles are mostly equatorial (Fig 3a), and as a result the polar axis is more subject to tradeoffs than is the equatorial one. The maximum $\chi$ at the 68% level is smaller, partly because *N* is larger but also each image constrains a different part of parameter space. Importantly, the estimated uncertainties in *a* and *c* are smaller than the "true" oblateness (*a-c*=1 pixel) which implies that an oblateness as small as 0.3% can in principle be detected. Solving for a spherical body using the same realizations results in a best-fit radius which is within 0.2 km (0.05 pixel) of the mean radius of the synthetic oblate body.

| Synthetic | N | $\chi/\chi_{min}$ (68%) | $|\Delta a|$ (68%) pix | $|\Delta c|$ (68%) | *(a-c)/a* (68%) |
|---|---|---|---|---|---|
| h=1 | 2059 | 1.0090 | 0.05 | 0.10 | 0.26-0.37% |
| h=1.5 | 2059 | 1.0088 | 0.10 | 0.20 | 0.21-0.37% |
| h=2.0 | 2059 | 1.0094 | 0.10 | 0.25 | 0.21-0.42% |

**Table B2**. As for Table B1 but using four projected profiles. Weighting of each profile is inversely proportional to image resolution. The errors in *a* and *c* are quoted in pixels for the highest-resolution image (3.65 km/pixel). The final column gives the (68%, two-sided) range of flattening determined. The true value is 0.3%.

### B.4 Effect of centroid uncertainty

A potential additional source of error is that the projection of the picks onto the sphere assumes that there are no uncertainties in the centroid associated with each profile ($x_0, y_0$). However, determination of the centroid is subject to uncertainties, and if the centroid is in error, the shape determined from combined projected limb picks will also be in error. We quantified this effect as follows.

Illuminated limb picks ($x_i, y_i$) from a particular image were read in. A "cloned" set of projected picks was then produced by displacing the image centroid by $\Delta x, \Delta y$ from the best-fit centroid (Table 3). The radial distance and projected location of each pick was then calculated, with the mean radius kept at the best-fit value. This process was repeated 30 times, producing 30 different sets of projected picks, each having a different centroid error. Based on Section B.2 the errors $\Delta x$ and $\Delta y$ were assumed to be uniformly distributed around zero with a maximum range of +/- 0.2 and +/- 0.5 pixels respectively.

Cloned projected picks were produced for visits 70p, 71p, 72p and 75p. For each set of cloned images, the best-fit oblate spheroid shape was determined as in Section 4. From our 30 clones, the standard deviation in the derived values of $a, c$ and $R$ were 0.68 km, 1.17 km and 0.68 km, respectively. These represent a measure of the uncertainty introduced by errors in the centroid.

### B.5 Summary

Based on our synthetics, for single images, the 2-$\sigma$ error ellipse is taken to be determined by $\chi = 1.044 \chi_{min}$ (Section B.2). Here it is assumed that noise and rough topography are the only contributors to the uncertainty.

When fitting multiple, projected sets of limb picks the situation is more complicated. Noise and rough topography contribute a 1-$\sigma$ uncertainty of 0.3 and 0.6 km to $a$ and $c$, respectively, and 0.2 km to $R$ if sphericity is assumed (Section B.3). Uncertainties in the centroid location contribute an additional uncertainty of 0.7 km, 1.2 km and 0.7 km (Section B.4).

For a spherical body, the combined 1-$\sigma$ uncertainty in $R$ is (conservatively) 1 km, while for $a$ and $c$ the combined uncertainties are 1.0 and 1.8 km. Based on Section B.3, a misfit of $1.01\chi_{min}$ yields a radius uncertainty of about 0.05 pixel (0.2 km) at the 1-$\sigma$ level. A 1 km radius uncertainty therefore implies a 1-$\sigma$ misfit $\chi \approx 1.05\chi_{min}$. Accordingly, when carrying out fitting to projected multiple sets of picks (Table 6), we take the 2-σ uncertainty to be $\chi/\chi_{min}$=1.1. This does not include contributions from image smear or camera distortion.

Appendix C: Radius determination via solar occultation using the Fine Sun Sensor

*New Horizons*' Fine Sun Sensor (FSS) is a component of the spacecraft's Guidance & Control system which identifies the position angle of the Sun. It is roughly co-aligned with the spacecraft antenna direction (+Y axis), and measures solar position away from this angle [Kagan 2003]. Typically, the FSS is used for spacecraft pointing determination, but we here analyze the FSS data to constrain the Pluto and Charon radii from occultations.

The FSS position uses a set of photometers arranged below a sandwiched pair of fine angular grating masks to measure the Sun's position in the sky, across a field-of-view of roughly 20°x20°. The instrument measures two independent quantities. The first is a 12-bit value which encodes an angular position for the Sun, at a stated resolution and accuracy of 0.05° and 0.004°, respectively. The second is a one-bit `Data Good' flag, indicating whether an illumination level above a specified threshold was received on a separate detector, adjacent and co-aligned with the photometers. Both data quantities are output at 25 Hz.

We use the FSS data to measure the ingress and egress times for the Pluto and Charon occultations. We do this by comparing the measured boresight-to-sun solar angle reported by the FSS, and comparing this to the boresight-to-sun angle measured by NH's star trackers. Where we see a deviation between these two angles, we interpret it as being caused by the partial occultation of Sun, which changes the position of its centroid.

Figure C1 shows the solar angles measured by the FSS (sun sensor) and FSW (star tracker) positions. Throughout the encounter, these curves track each other to typically <0.001°, except during the occultations. At ingress, the deviation reaches ~0.005° (slightly less than the solar radius of 0.008°) before the FSS signal diverges and goes to zero, over the course of roughly 15 seconds. The Data Good flag cuts off roughly 1 sec before the FSS data go invalid. At egress (shown), the FSS signal returns but deviates from FSW, until the last contact at which point the signals again track each other well. The Charon occultation is shown for clarity because it is longer than that for Pluto. Ingress and egress are similar and there are no significant asymmetries.

From the FSS results we measure occultation lengths of 663.0 sec (Pluto) and

220.3 sec (Charon), from ingress midpoint to egress midpoint. For Charon the ingress and egress times can be measured to a precision of 0.1 sec (3-σ) while for Pluto the signal is noisier and the uncertainty is 0.3 sec. We can then apply the geometry model of the New Horizons mission's SPICE kernel set *od122* to infer the chord lengths, and derived body radii. By measuring the offset between our measured occultation times and the SPICE-derived times, we can also determine a correction $\delta t$ to the along-track spacecraft position.

Our results are shown in Table C1. For Pluto we derive a radius of 1189±2 km, essentially identical to the LORRI-derived radius. This is an 'optical radius' which may include some atmospheric effects which were not seen in low-phase approach imaging. Our derived value for Charon's radius is 619.0±0.5km, larger than the LORRI-derived value of 606.0 ± 1.0 km. This difference is larger than typical topographic variations on Charon. However, the grazing geometry of the Charon event magnifies magnifies small distances such that the local topography or a small offset in Charon's radius or position translates into a larger change in the chord length.

For both Pluto and Charon, we found an offset of $\delta t$ = 0.54 sec to fit well, with an estimated uncertainty of 0.1 sec for Charon and 0.5 sec for Pluto. The fact that the same value fits both occultations is consistent with this being a correction to the along-track path of *New Horizons*, not to the individual orbits of Pluto and Charon. A positive value for $\delta t$ implies that *New Horizons* arrives later than specified in the trajectory kernel.

|  | Pluto | | Charon | |
| --- | --- | --- | --- | --- |
|  | **Predict** | **Measured** | **Predict** | **Measured** |
| First Contact | 12:44:19.0 | 12:44:14.8 ± 0:2 | 14:14:10.3 | 14:14:06.9 ± 0:1 |
| Midpoint Ingress | 12:44:20.9 | 12:44:16.7 ± 0:2 | 14:14:17.3 | 14:14:13.9 ± 0:1 |
| Totality Start | 12:44:22.8 | 12:44:18.6 ± 0:2 | 14:14:24.6 | 14:14:21.2 ± 0:1 |
| Totality geometrical midpoint |  | 12:49:50.5 ± 0:2 |  | 14:16:06.9 ± 0:1 |
| Totality End | 12:55:18.9 | 12:55:17.5 ± 0:2 | 14:17:50.1 | 14:17:47.0 ± 0:1 |
| Midpoint Egress | 12:55:21.1 | 12:55:19.8 ± 0:2 | 14:17:57.6 | 14:17:54.5 ± 0:1 |
| Last Contact | 12:55:23.4 | 12:55:22.0 ± 0:2 | 14:18:04.8 | 14:18:01.7 ± 0:1 |
| Duration (Totality)(s) | 656.1 | 658.8 | 192.6 | 205.5 |
| Duration (Mid-Mid) (s) | 660.2 | 663.0 | 206.2 | 220.3 |
| Duration (First-Last)(s) | 664.4 | 667.2 | 222.6 | 234.5 |
| Velocity (km/s) | 3.59 | 3.59 | 3.55 | 3.55 |

| Chord Length (km) | 2357.9 | 2367.9 | 678.2 | 730.5 |
|---|---|---|---|---|
| Body Radius(km) | 1184 | 1189 ± 2 | 603.5 | 619.0 ± 0.5 |
| Offset $\delta t$ (s) | 0 | 0.54 ± 0.5 | 0 | 0.54 ± 0.1 |
| Ingress midpoint lon, lat (RHR) | -161.1°, -20.1° | | 53.2°, 11.0° | |
| Egress midpoint lon, lat (RHR) | 9.9°, 11.8° | | 131.8°, 34.1° | |
| Distance to body (km) | 52,527 | | 115,158 | |
| | | | | |

**Table C1** Times for the Pluto and Charon FSS occultation events. Predict times are calculated using SPICE and assuming mission-derived body radii. The times are calculated geometrically and ignore any atmospheric effects. 'First contact', 'midpoint' and 'totality' are defined to correspond to flux levels of 100%, 50% and 0%. Meausred times are derived using the same trajectory, but the body radii and NH's along-track time offset $\delta t$ are free parameters, and are adjusted to fit the FSS data. The 'first contact' and 'last contact' are directly measured from the FSS data, while the other times are based on the derived body radii. The distances are center-center at the geometric occultation middle. Error bars for radii and $\delta t$ are 3σ.

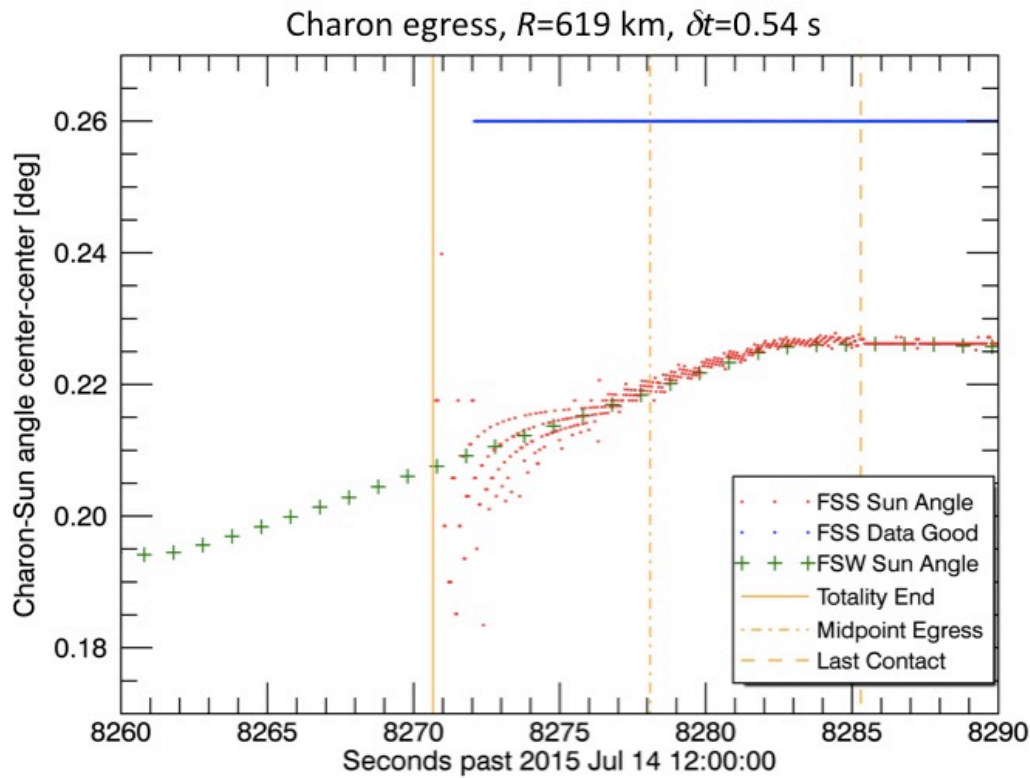

**Figure C1** The angular position of the Sun as measured by the FSS instrument (red dots) and Star Trackers (green +) are shown for the egress after the Charon occultation. During the occultation the FSS reports no valid data, and after egress the two values track closely. During egress from totality to last contact, the FSS data show enhanced scatter, perhaps due to diffraction effects within the instrument's slit mask. The overall slope of the curves is due to the spacecraft's rotation, and the change in slope near 8282 sec is due to a thruster firing. The orange lines show the occultation egress times, as predicted by the *od122* kernel set and shifted in time by our derived offset of $\delta t$ = 0.54 sec.

Appendix D: The ALICE instrument and occultation geometry

The ALICE instrument onboard *New Horizons* probes a wavelength region between 52 – 187 nm [Slater *et al.* 2005], and its field of view consists of a 2$^o$ x 2$^o$ box at the top of a narrow (0.1$^o$) slot. During observations of the sun, the solar occultation channel (or SOCC) is used, and sunlight is reflected off a small pickoff mirror to the rest of the primary optics. The photons are collected onto a 32 row by 1024 column imaging array, with spectral information contained in the *x*-direction and spatial information in the *y*-direction. The data shown here was acquired after closest approach while passing into Charon's shadow during observation of a solar occultation [Stern *et al.* 2008].

Before analysis of the raw spectra, the data is first corrected for several instrumental effects. Dead time in the readout of the pixel list information is corrected at the native time resolution of 4 milliseconds, before later binning to a time resolution of 1 second. Stim pixels are used to correct for temperature effects on the detector, as resistivity affects mapping of physical location on the array to a given pixel number.

During the solar occultation observation, the majority of the photons fell within a few rows on the detector, and so the analysis presented here sums over rows 19-22 to create a single 2-D spectral image of instrument counts in wavelength and time dimensions. Reconstructed spacecraft geometry is then used to convert time values at 1 second resolution into values of tangent radius, where tangent radius is the projected distance of the center of the Sun from the center of Charon as seen by *New Horizons*. These tangent radii values are useful for atmospheric studies and can also be used to determine the length of the occultation chord across the disk of the planet.

The solar occultation by Charon occurs relatively quickly, as the projected velocity of the spacecraft in the viewing plane is 3.55 km/s. However, the time resolution of the measured count rate allows for modeling of the slope in transmission seen during ingress and egress. This slope is accounted for by the finite size of the Sun as seen by ALICE, which has an effective radius of about 16 km given the relative distances of the Sun-Charon-spacecraft. This value has been calculated using Equation D1 below.

$$R_{eff} = \frac{R_{sun} D_{Charon}}{D_{sun}} \qquad (D1)$$

In this calculation, $R_{sun}$ is $6.955 \times 10^5$ km, $D_{sun}$ is 32.9 AU ($4.923 \times 10^9$ km) and $D_{Charon}$ is approximately 113,577 km at limb contact during ingress and 116,730 km at limb contact during egress. This results in effective solar radii of 16.05 km and 16.49 km, respectively. In order to extract the radius of Charon, these values are then used in a forward model to match the detected slope in transmission as a function of tangent radius. This forward model is essentially a calculation of the percent overlap between a circle with radius $R_{eff}$ and a 'knife-edge' that occults the area of the circle over time, represented mathematically as Equation D2:

$$F = \frac{R_{eff}^2 \arccos\left(\frac{D_{limb}}{R_{eff}}\right) - D_{limb}\sqrt{R_{eff}^2 - D_{limb}^2}}{\pi R_{eff}^2} \qquad (D2)$$

In this equation, $F$ is the fraction of disk that is occulted given $R_{eff}$, the effective solar radius, and $D_{limb}$, the distance of the center of the sun from the edge of the limb. Combination of this term with the tangent radius (distance of the center of the sun from Charon center) then allows for estimation of Charon's radius, or the distance from the edge of the limb to the center of Charon. The results of a forward model for both ingress and egress are shown in Figures D1 and D2 for the total transmission over the entire ALICE bandpass, as well as transmission for diagnostic regions from 140-190 nm, 100-140 nm, and 50-100 nm. Note that differences in noise correspond with total count rate for each region, as there is less solar flux at the shorter wavelengths.

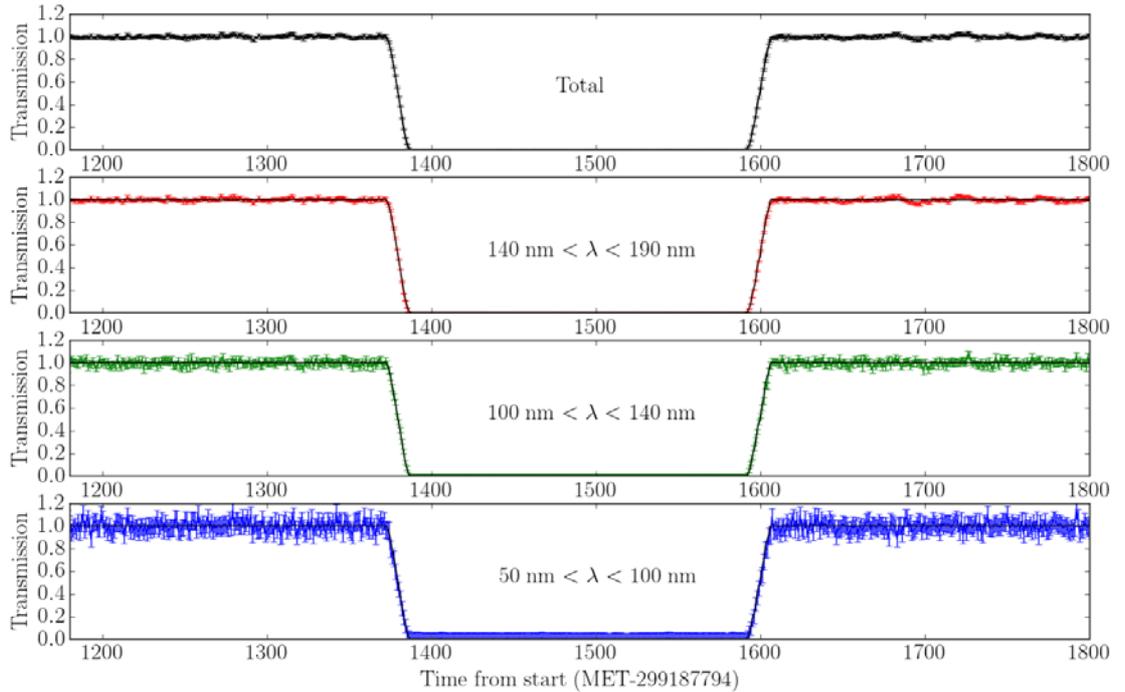

**Figure D1.** Transmission as a function of time at several selected wavelength regions. Best fit forward model is shown as black solid line.

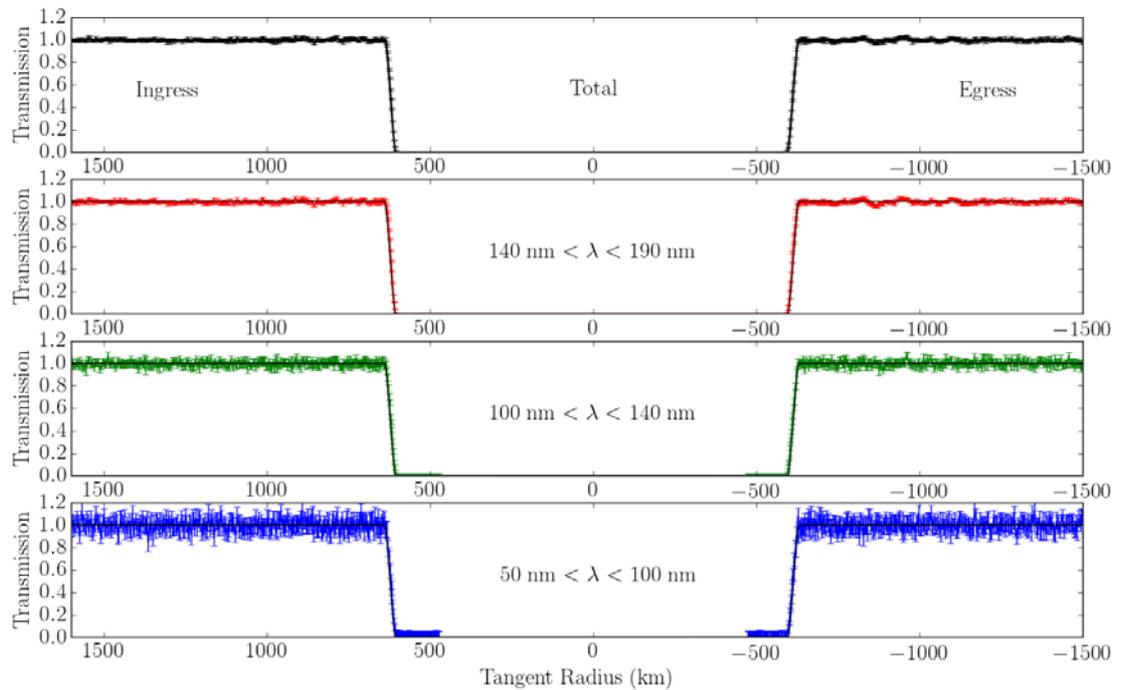

**Figure D2**. Transmission as a function of tangent radius at several selected wavelength regions. Best fit forward model is shown as black solid line

The time of mid-ingress (estimated at the half-light value) is 1379.69 s after the start of the observation, and the time of mid-egress is 1600.05 s after start, resulting in a chord length of 220.36 s, or equivalently (given the projected speed of 3.55 km/s) 782 km.

The best estimates for the radius of Charon at entry and exit points for the data shown here are 619 km (ingress) and 610 km (egress) (see Table D1 below). However, while the timing of ingress and egress are extremely well-constrained, the exact value of tangent radii during the occultation is less well-known. Based on currently known spacecraft ephemeris, the propagated 1-sigma uncertainty of tangent radius at entry and exit is +/- 7 km. This likely accounts for the relatively high and asymmetric best-fit radii.

Estimating the timing for Pluto is more difficult, since the atmosphere (and the haze in particular) absorbs so much of the UV. Estimating the time of totality and then working backwards, using the fact that during the Pluto occultation the projected size of the sun has a radius of ~7 km during ingress and ~8 km during egress, yields a radius of 1191 km with an uncertainty of a few km.

|                              | Pluto       | Charon        |
|------------------------------|-------------|---------------|
| First Contact                | -           | 14:14:08.2    |
| Midpoint Ingress             | -           | 14:14:15.7    |
| Totality Start               | 12:44:21    | 14:14:23.2    |
| Totality End                 | 12:55:21    | 14:17:48.9    |
| Midpoint Egress              | -           | 14:17:56.1    |
| Last Contact                 | -           | 14:18:03.3    |
| Duration (Midpt to Midpt, s) | 664         | 220.4         |
| Velocity (km/s)              | 3.59        | 3.55          |
| Chord (km)                   | 2383        | 782           |
| Body Radius (km)             | $1191 \pm 3$ | $(619, 610) \pm 7$ |

**Table D1.** Summary of ALICE occultation results. For Charon, radius results for ingress and egress are reported separately.

Appendix E: LORRI image distortion

One description of the LORRI distortion is provided by the Simple Imaging Polynomial distortion model (SIP), which is prevalent in the astronomy community and is described in Shupe *et al.* [2005]. The SIP polynomial parameters given below are documented in Steffl et al. [2016] and were derived from LORRI calibration data reported in Owen and O'Connell [2011].

For a pixel observed at ($x,y$) relative to the centre of the image at ($x_0,y_0$), we define $u=x-x_0$ and $v=y-y_0$. Then for a third-order polynomial the horizontal and vertical distortion $\delta x$ and $\delta y$ are given by

$$\delta x = A_{20}u^2 + A_{02}v^2 + A_{11}uv + A_{21}u^2v + A_{12}uv^2 + A_{30}u^3 + A_{03}v^3$$

$$\delta y = B_{20}u^2 + B_{02}v^2 + B_{11}uv + B_{21}u^2v + B_{12}uv^2 + B_{30}u^3 + B_{03}v^3$$

where $A_{ij}$ and $B_{ij}$ are polynomial coefficients tabulated below. The original (pre-distorted) limb location ($x',y'$) is then given by $x'=x-\delta x$, $y'=y-\delta y$.

| $A_{30}$ | -4.5683524653106E-09 | $B_{30}$ | -4.8263374371619E-16 |
| $A_{21}$ |  3.6773993329229E-13 | $B_{21}$ | -4.5505047160943E-09 |
| $A_{12}$ | -4.5506608174421E-09 | $B_{12}$ |  3.6773991492864E-13 |
| $A_{03}$ | -4.8263827227450E-16 | $B_{03}$ | -4.5685088916275E-09 |
| $A_{20}$ |  3.7132883452972E-07 | $B_{20}$ | -2.5764535470748E-10 |
| $A_{11}$ |  2.4489911491959E-07 | $B_{11}$ |  3.7063022991452E-07 |
| $A_{02}$ | -3.8995992016687E-10 | $B_{02}$ |  2.4536068067188E-07 |

**Table E1.** SIP coefficients describing LORRI distortion (see text).